\documentclass[%
reprint,
superscriptaddress,
 amsmath,amssymb,
 aps,
prx,
longbibliography
]{revtex4-2}

\usepackage{braket}
\usepackage[english]{babel}
\usepackage{graphicx}% Include figure files
\usepackage{dcolumn}% Align table columns on decimal point
\usepackage{bm}% bold math
\usepackage{algorithm,algpseudocode}
\usepackage{pifont}
\usepackage{xcolor}
\definecolor{darkgreen}{RGB}{40,150,100}
\usepackage[colorlinks=true,
    linkcolor=red,     % color for internal links (like section refs)
    citecolor=darkgreen,   % color for citations
    urlcolor=blue      % color for URLs
]{hyperref}

\usepackage{tikz}
\newcommand*\circnum[1]{%
  \tikz[baseline=(char.base),scale=0.85]{
    \node[draw,circle,inner sep=0.5pt] (char) {#1};}}
\makeatletter
\renewcommand{\fnum@figure}{Fig.~\thefigure}
\makeatother

\begin{document}

\preprint{chirp gates}

\title{Frequency- and Amplitude-Modulated Gates for Universal Quantum Control}% Force line breaks with \\

\def\EECSaffil{Department of Electrical Engineering and Computer Science, Massachusetts Institute of Technology, Cambridge, MA 02139, USA}
\def\RLEaffil{Research Laboratory of Electronics, Massachusetts Institute of Technology, Cambridge, MA 02139, USA}
\def\LLaffil{MIT Lincoln Laboratory, Lexington, MA 02421, USA}
\def\Physaffil{Department of Physics, Massachusetts Institute of Technology, Cambridge, MA 02139, USA}
\def\Googaffil{Google Quantum AI, Santa Barbara, CA 93111, USA}

\author{Qi~Ding}
\email{qding@mit.edu}
\affiliation{\EECSaffil}
\affiliation{\RLEaffil}

\author{Shoumik~Chowdhury}
\affiliation{\EECSaffil}
\affiliation{\RLEaffil}

\author{Agustin~Di~Paolo}
\affiliation{\Googaffil}

\author{Réouven Assouly}
\affiliation{\RLEaffil}

\author{Alan~V.~Oppenheim}
\affiliation{\EECSaffil} 
\affiliation{\RLEaffil}

\author{Jeffrey~A.~Grover}
\affiliation{\RLEaffil}

\author{William~D.~Oliver}
\email{william.oliver@mit.edu}
\affiliation{\EECSaffil} 
\affiliation{\RLEaffil} 
\affiliation{\Physaffil}

\date{\today}

\begin{abstract}
Achieving high-fidelity single- and two-qubit gates is essential for executing arbitrary digital quantum algorithms and for building error-corrected quantum computers. We propose a theoretical framework for implementing quantum gates using frequency- and amplitude-modulated microwave control, which extends conventional amplitude modulation by introducing frequency modulation as an additional degree of control. Our approach operates on fixed-frequency qubits, converting the need for qubit frequency tunability into drive frequency modulation. Using Floquet theory, we analyze and design these drives for optimal fidelity within specified criteria. Our framework spans adiabatic to nonadiabatic gates within the Floquet framework, ensuring broad applicability across gate types and control schemes. Using typical transmon qubit parameters in numerical simulations, we demonstrate a universal gate set—including the X, Hadamard, phase, and CZ gates—with control error well below 0.1\% and gate times of 25--40 ns for single-qubit operations and 125--135 ns for two-qubit operations. Furthermore, we show an always-on CZ gate tailored for driven qubits, which has gate times of 80--90 ns.
\end{abstract}

\maketitle
\section{Introduction} \label{sec:introduction}
High-fidelity single- and two-qubit gates are required to implement arbitrary digital quantum algorithms and for realizing error-corrected quantum computers~\cite{zajac_Resonantly_2018,wright_Benchmarking_2019,he_Twoqubit_2019,krantz_Quantum_2019}. Among the available hardware platforms, superconducting qubits stand out as a leading candidate, featuring sufficient gate fidelity to enable prototype demonstrations of quantum error detection~\cite{arute_Quantum_2019,ding_HighFidelity_2023,zhang_Tunable_2024,rower_Suppressing_2024,kjaergaard_Superconducting_2020,sivak_Realtime_2023,kim_Evidence_2023,acharya_Suppressing_2023,acharya_Quantum_2025,marxer_999_2025}. Despite significant advances in superconducting qubits, continued efforts aim to further improve gate fidelities, as higher fidelities directly translate to deeper executable circuits and reduced overhead requirements to achieve a desired error rate via error-correction protocols.

In transmon-based devices, quantum gates, especially two-qubit entangling gates, can be broadly classified as either microwave activated or baseband flux controlled. Microwave-activated gates~\cite{chow_Simple_2011,chow_Microwaveactivated_2013,mckay_Universal_2016,krinner_Demonstration_2020,wei_Hamiltonian_2022} are typically implemented with fixed-frequency qubits, which exhibit longer coherence times and reduce control overhead, but are limited by potential frequency crowding issues as system size increases. Baseband-flux-controlled gates~\cite{neeley_Generation_2010,chen_Qubit_2014,barends_Diabatic_2019,campbell_Universal_2020, sung_Realization_2021,ding_Pulse_2025,marxer_999_2025,an2025zzfreetwotransmonczgate}, implemented with tunable qubits and baseband magnetic flux pulses, enable faster gate execution with generally higher fidelity and mitigate frequency-collision issues, but increase hardware and calibration complexity as well as sensitivity to flux noise. Most of these gates that use analytical waveforms are implemented with a fixed carrier frequency (microwave-activated) or no carrier frequency (baseband-flux-controlled), while the amplitude envelope of the control pulse is deliberately shaped to achieve high fidelity. Other approaches exist where numerical methods such as quantum optimal control~\cite{krotov_Global_1993,khaneja_Optimal_2005,chen_Accelerating_2023,boscain_Introduction_2021,hyyppa_Reducing_2024,lin_Timeoptimal_2025} or learning-based approaches~\cite{niu_Universal_2019,palittapongarnpim_Learning_2017,zhang_When_2019,baum_Experimental_2021,sivak_ModelFree_2022,reuer_Realizing_2023,ding_HighFidelity_2023} are used to directly design or optimize the control pulses. 
% Fast nonadiabatic control [cite our many papers] is another method to implement gates.

The concept of frequency modulation for CZ gates in superconducting qubits was proposed in Ref.~\cite{paolo_Extensible_2022}, where changing the drive frequency in frequency-modulated gates can be viewed as an analog of changing the qubit frequency in baseband-flux-controlled gates. In this work, we generalize this idea to realize a high-fidelity universal gate set. We present a theoretical framework that extends conventional amplitude modulation by incorporating frequency modulation as an additional control dimension, enabling the design of both adiabatic and nonadiabatic gates. Floquet theory is used as the foundation for constructing this framework. Using numerical simulations with typical transmon parameters, we construct a universal gate set—including the X, Hadamard, phase, and CZ gates—achieving control error below 0.1\% with gate durations of 25--40 ns for single-qubit gates and 80--135 ns for two-qubit gates. In addition, we present an always-on version of the CZ gate for driven qubits, which substantially shortens the gate time relative to the standard CZ gate. Our simulations only account for dynamical control errors and there is no decoherence. These findings indicate that microwave control schemes based on both amplitude and frequency modulation provide a promising approach for implementing high-fidelity quantum gates, affording greater flexibility in gate design.

The paper is organized as follows. In Section~\ref{sec:framework}, we introduce the general theoretical framework and provide a qualitative overview of the gate principles for both the adiabatic and nonadiabatic gates. We also present an explicit pulse-optimization strategy for designing both types of gates. Section~\ref{sec:adiabatic_gates} presents examples of adiabatic gates, including the CZ and Z gates, while Section~\ref{sec:nonadiabatic_gates} discusses examples of nonadiabatic gates, including the X and Hadamard gates. Finally, we conclude and discuss the directions of future extensions in Section~\ref{sec:conclusion}.

\section{General Theoretic Framework}\label{sec:framework}
In this section, we present the fundamental principles underlying frequency- and amplitude-modulated gates within a standard qubit-coupler-qubit transmon architectural primitive. We first provide a qualitative description of the theoretical framework for analyzing gate dynamics and summarize the key aspects of Floquet theory that form the foundation of this framework. Finally, we outline a quantitative optimization strategy for control parameters that achieves both high-fidelity and fast gates.

\begin{figure*}[htbp]
    \centering
    \includegraphics[width=1.\linewidth]{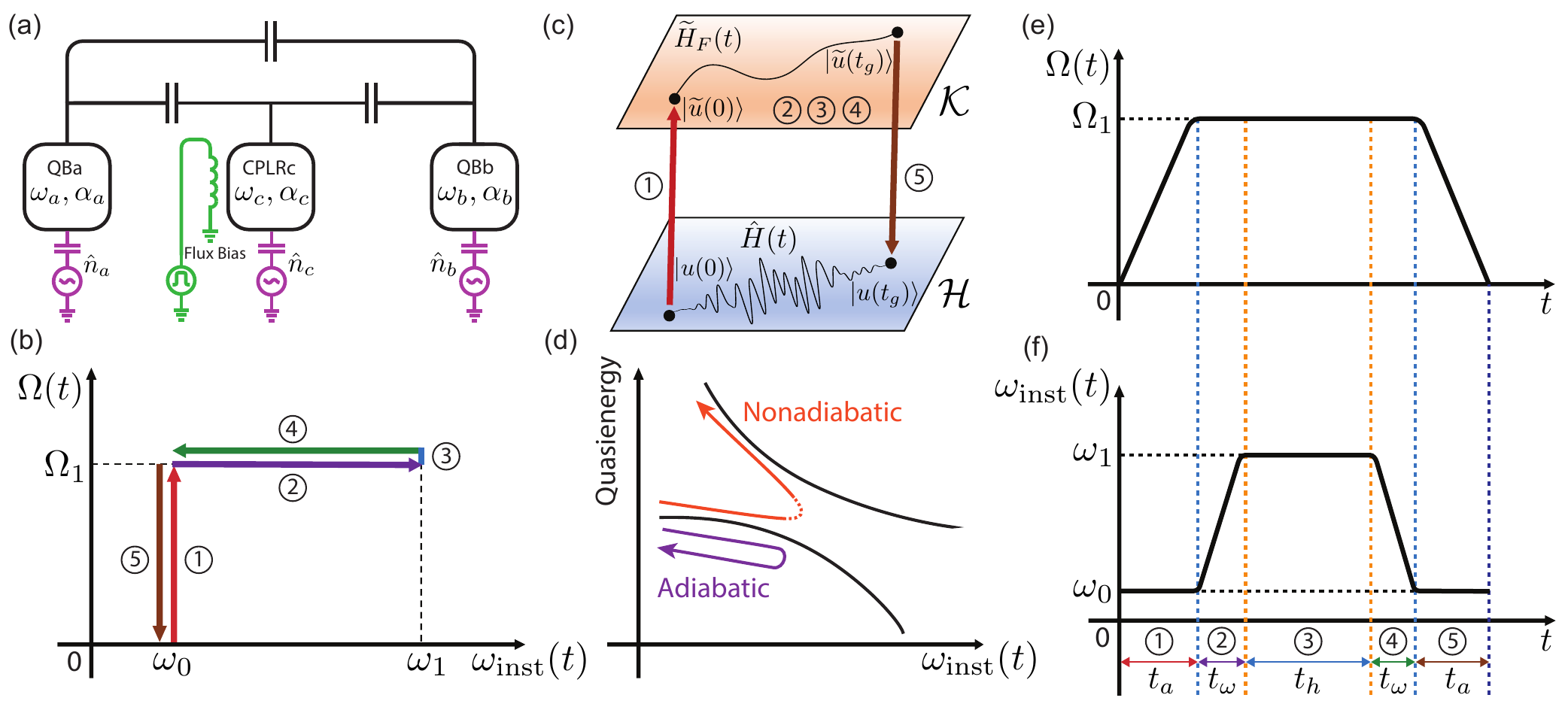}
    \caption{Overview of the gate protocol. (a) Qubit-coupler-qubit transmon-based circuit with two fixed-frequency qubits (QBa, QBb) and a flux-tunable coupler (CPLRc). Each qubit (coupler) has a microwave charge drive line. Proposed gates are implemented using microwave control only. (b) The staged gate protocol manifested in the $\omega_\text{inst}(t)-\Omega(t)$ parameter space. In stages \protect\circnum{1}, \protect\circnum{5}, we turn on/off the drive amplitude while keeping $\omega_\text{inst}(t)=\omega_0$. In stages \protect\circnum{2}, \protect\circnum{4}, we modulate the drive frequency while keeping $\Omega(t)=\Omega_1$. In stage \protect\circnum{3}, we hold at $\Omega(t)=\Omega_1, \omega_\text{inst}(t)=\omega_1$. (c) Mapping between the original Hilbert space $\mathcal{H}$ and the extended Hilbert space $\mathcal{K}$, where the example system is driven by a frequency-modulated microwave pulse. Stage \protect\circnum{1}: turn on the drive, mapping from $\mathcal{H}$ to $\mathcal{K}$. Stages \protect\circnum{2}, \protect\circnum{3}, \protect\circnum{4}: state evolution can be best analyzed and designed in $\mathcal{K}$ where the dynamics are simpler. Stage \protect\circnum{5}: turn off the drive, mapping from $\mathcal{K}$ back to $\mathcal{H}$. (d) High-level illustrations of adiabatic gates (purple trajectory, e.g., Z and CZ gates) and nonadiabatic gates (orange trajectory, e.g., X gate). (e)(f) Descriptive profiles of $\Omega(t)$ and $\omega_\text{inst}(t)$, and the segmentation with time parameters corresponding to the staged gate protocol.}
    \label{fig:general_framework}
\end{figure*}

\subsection{Preliminaries}
We consider the qubit-coupler-qubit transmon-based circuit illustrated in Fig.~\ref{fig:general_framework}(a). The qubits (QBa, QBb) are fixed-frequency, while the coupler (CPLRc) is frequency-tunable by an external magnetic flux to frequencies suitable for implementing the desired gates. The gate schemes we propose rely exclusively on all-microwave drives, avoiding the need for baseband, fast-flux pulses and their associated challenges. Single-qubit gates are implemented by driving the target qubit through its charge line. For two-qubit CZ gates, we instead drive the coupler through its charge line, inducing a nonzero ZZ interaction. We model each qubit and coupler as a Kerr nonlinear oscillator, a common model for anharmonic multi-level qubit systems such as the transmon~\cite{koch_Chargeinsensitive_2007}, with Hamiltonian ($\hbar:=1$):
\begin{equation}
    \hat{H}_j = \omega_j \hat{a}_j^\dagger \hat{a}_j + \frac{\alpha_j}{2}\hat{a}_j^\dagger\hat{a}_j^\dagger \hat{a}_j
    \hat{a}_j,
\end{equation}
where $\omega_j$ and $\alpha_j$ are the frequency and anharmonicity of mode $j$, respectively, and $\hat{a}^\dagger_j$ ($\hat{a}_j$) are the corresponding raising (lowering) operators. Here, $\ j\in\{a,b,c\}$ denotes QBa, QBb, and CPLRc, respectively. There is capacitive coupling between each pair of modes, and the interaction terms between the modes are of the form
\begin{equation}
    \hat{H}_{jk} = -g_{jk}(\hat{a}_j - \hat{a}^\dagger_j)(\hat{a}_k - \hat{a}^\dagger_k),
\end{equation}
where $g_{jk}$ is the coupling rate between mode $j$ and $k$.
The system Hamiltonian is then taken to be
\begin{equation}
    \hat{H}_0 = \sum_{j} \hat{H}_j + \sum_{j\neq k} \hat{H}_{jk} .
\end{equation}
Note that by $j\neq k$, we mean $(j,k)\in\{(a,b),(b,c),(c,a)\}$. The time-dependent drive Hamiltonian is modeled by
\begin{equation}
    \hat{H}_d(t) = \Omega(t)\cos[\theta(t)]\hat{n}_j ,
\end{equation}
where $\hat{n}_j = i(\hat{a}^\dagger_j- \hat{a}_j)$ represents the charge operator of the mode $j$ to which the drive is applied. Here, $\Omega(t)\geq0$ is the drive amplitude envelope and $\theta(t)$ is the drive angle, which is related to the instantaneous drive frequency $\omega_{\rm inst}(t)$ via $\omega_{\rm inst}(t) = \dot{\theta}(t)$. We therefore have $\theta(t)=\int_0^t\omega_{\rm inst}(\tau)\text{d}\tau$. In this work, we focus on designing $\omega_\text{inst}(t)$, from which we can easily obtain $\theta(t)$. For example, for fixed-frequency drives with $\omega_{\rm inst}(t)=\omega_d$, the drive angle is $\theta(t)=\omega_dt$. Throughout this paper, we refer to the design of $\Omega(t)$ as \emph{amplitude modulation} and the design of $\omega_\text{inst}(t)$ as \emph{frequency modulation}~\cite{oppenheim1997signals}. In simulations, we report control error $1-\mathcal{F}$, where $\mathcal{F}$ is the coherent average gate fidelity defined in Appendix~\ref{app:definitions}.

\subsection{Qualitative picture of the gate-design principles}
\subsubsection{Floquet theory overview}
We first briefly summarize the extended Hilbert space representation of generalized Floquet theory~\cite{floquet_Equations_1883,sambe_Steady_1973,shirley_Solution_1965,eckardt_Highfrequency_2015,rudner_Floquet_2020,silveri_Quantum_2017a,howland_Stationary_1974,chu1985advances,bialynicki-birula_Quantum_1976,guerin_Complete_1997,guerin_Control_2003} and introduce the basic notations used throughout this work (see Appendix~\ref{app:floquet} for more details). Consider the system Hamiltonian $\hat{H}(t)=\hat{H}_0 + \hat{H}_d(t)$, where the drive Hamiltonian is $\hat{H}_d= \Omega(t)\cos[\theta(t)]\hat{n}_j$ in the original Hilbert space $\mathcal{H}$. We promote $\theta(t)$ to a $2\pi$-periodic quantum degree of freedom $\hat{\vartheta}$ whose conjugate variable is $\hat{m}=-i\partial_{\vartheta}$, satisfying $[\hat{\vartheta},\hat{m}]=i$. Then, we can write the Floquet Hamiltonian in the extended Hilbert space $\mathcal{K}$ as:
\begin{equation}\label{eq:floquet_H_compact}
    \widetilde{H}_F(t) = \hat{H}_0 +\Omega(t)\cos(\hat{\vartheta})\hat{n}_j + \omega_\text{inst}(t) \hat{m},
\end{equation}
where $\cos(\hat\vartheta) = \frac{1}{2}\sum_m \ket{m+1}\!\bra{m}{\rm\  +\  h.c.}$ and $\hat{m}=\sum_m m\ket{m}\!\bra{m}$ in the $\hat{m}$-basis, with $m\in\mathbb{Z}$. Here, $m$ can be interpreted as the number of photons subtracted or added from the drive field when a transition is driven. We denote the instantaneous eigenenergies and eigenstates of $\widetilde{H}_F(t)$ by $\varepsilon_{m,\alpha}(t)$ and $\ket{\widetilde{u}_{m,\alpha}(t)}$, which are also referred to as quasienergies and Floquet modes, respectively. A more mathematical construction of the extended Hilbert space can be found in Appendix~\ref{app:floquet}.
When the drive amplitude $\Omega$ is zero, the Floquet modes $\ket{\widetilde{u}_{m,\alpha}} =\ket{m}\otimes\ket{\alpha}$ are product states with eigenenergies $\varepsilon_{m,\alpha} = E_\alpha+m\omega_\text{inst}$, with $E_\alpha$ being the energy of the (undriven) qubit basis state $\ket{\alpha}$ in $\mathcal{H}$.
Then, at nonzero drive amplitudes $\Omega > 0$, the Floquet modes $\ket{\widetilde{u}_{m,\alpha}} =\ket{m,\alpha}$ are no longer product states due to the hybridization between the qubit and the driving field. The quasienergies of different Floquet modes vary parametrically as a function of $\Omega(t)$ and $\omega_\text{inst}(t)$, and therefore can become close and form avoided crossings. We leverage such avoided crossings to design gates in our protocol.

\subsubsection{Staged gate protocol}
The proposed gate protocol is divided into five stages in the $\omega_\text{inst}(t)-\Omega(t)$ parameter space shown in Fig.~\ref{fig:general_framework}(b). Figure~\ref{fig:general_framework}(c) depicts the evolution during the five stages in $\mathcal{H}$ and $\mathcal{K}$. Figures~\ref{fig:general_framework}(e)--(f) show descriptive profiles of $\Omega(t)$ and $\omega_\text{inst}(t)$ corresponding to the five stages:

\begingroup
\vspace{1\baselineskip}
\leftskip=1em
\noindent \circnum{1} $t\in[0,t_a]$: Amplitude ramp of duration $t_a$ for $\Omega(t)$ to go from \( 0 \) to \( \Omega_1>0 \), while holding $\omega_\text{inst}(t)=\omega_0$.  
\par
\endgroup
\vspace{0.25\baselineskip}
\begingroup
\leftskip=1em
\noindent \circnum{2} $t\in[t_a, t_a+t_\omega]$: Frequency chirp of duration $t_\omega$ for $\omega_\text{inst}(t)$ to go from \( \omega_0 \) to \( \omega_1 \), while holding  \( \Omega(t)=\Omega_1 \). 
\par
\endgroup
\vspace{0.25\baselineskip}
\begingroup
\leftskip=1em
\noindent \circnum{3} $t\in[t_a+t_\omega, t_a+t_\omega+t_h]$: ``Hold'' stage of duration $t_h$ between stages \circnum{2} and \circnum{4}, where $\Omega(t) = \Omega_1$ and $\omega_{\text{inst}}(t) = \omega_1$.
\par
\endgroup
\vspace{0.25\baselineskip}
\begingroup
\leftskip=1em
\noindent \circnum{4} $t\in[t_a+t_\omega+t_h ,t_a+2t_\omega+t_h]$: Frequency chirp of duration $t_\omega$ for $\omega_\text{inst}(t)$ to go from \( \omega_1 \) to \( \omega_0 \), while holding \( \Omega(t)=\Omega_1 \).
\par
\endgroup
\vspace{0.25\baselineskip}
\begingroup
\leftskip=1em
\noindent \circnum{5} $t\in[t_a+2t_\omega+t_h, 2t_a+2t_\omega+t_h]$: Amplitude ramp of duration $t_a$ for $\Omega(t)$ to go from \( \Omega_1 \) to \( 0 \), while holding $\omega_\text{inst}(t)=\omega_0$.
\par
\endgroup
\vspace{1\baselineskip}
\noindent Since the gates we show are by design symmetric, we have introduced the time parameters $t_a, t_\omega, t_h$ as the duration for stages \circnum{1}/\circnum{5}, \circnum{2}/\circnum{4}, and \circnum{3}, respectively. The time parameters are labeled in Figs.~\ref{fig:general_framework}(e)--(f). Note that the relative ordering of $\omega_0$ and $\omega_1$ depends on the specific gate. This staged control strategy allows us to separate the roles of amplitude and frequency modulation: amplitude modulation ensures that the mapping between $\mathcal{H}$ and $\mathcal{K}$ is adiabatic, while frequency modulation activates desired interactions for the gate operation. While this is a restricted mode of operation compared with the full generality of simultaneous amplitude and frequency control that is possible, we choose to operate within these restrictions to make the potential experimental implementation simpler. Future studies may extend this work to incorporate general frequency and amplitude modulation through all stages \circnum{1}--\circnum{5}.

As depicted in Fig.~\ref{fig:general_framework}(c), the process of turning the drive on (stage \circnum{1}) maps from $\mathcal{H}$ to $\mathcal{K}$. Similarly, turning the drive off (stage \circnum{5}) is the inverse mapping. 
Provided that the magnitudes of $\dot{\Omega}(t)$ and $\dot{\omega}_\text{inst}(t)$ are much smaller than $\omega_\text{inst}(t)$, the main evolution (stages \circnum{2}, \circnum{3}, \circnum{4}) can be accurately and clearly described in $\mathcal{K}$, as illustrated in Fig.~\ref{fig:general_framework}(b) by the comparison between the evolution trajectories of $\ket{\widetilde{u}(t)}$ in $\mathcal{K}$ and $\ket{u(t)}$ in $\mathcal{H}$. The key idea is that the fast oscillatory components at frequency $\omega_\text{inst}(t)$ are absorbed into the instantaneously changing basis, and therefore a slowly varying $\omega_\text{inst}(t)$ is necessary to maintain the integrity of the desired evolution in $\mathcal{K}$. In the examples we show in Sections~\ref{sec:adiabatic_gates} and~\ref{sec:nonadiabatic_gates}, we choose $\max\{|\dot{\Omega}(t)|,|\dot{\omega}_\text{inst}(t)|\}/\min\{\omega_\text{inst}(t)\}<0.01$.

\subsubsection{Adiabatic and nonadiabatic gates}
We further classify the proposed gates into two categories: adiabatic and nonadiabatic, as illustrated in Fig.~\ref{fig:general_framework}(d). The adiabatic gate operates by closely following the instantaneous Floquet modes throughout the entire gate duration, whereas the nonadiabatic gate deliberately induces transitions between Floquet modes and exploits interference to complete the operation. We discuss the unitary propagator during the gates for both adiabatic and nonadiabatic gates leveraging the extended Hilbert space representation $\mathcal{K}$ in Appendix~\ref{app:unitary_propagator}. High-fidelity CZ and Z gates in Section~\ref{sec:adiabatic_gates} manifest the adiabatic approach, while the X and Hadamard gates in Section~\ref{sec:nonadiabatic_gates} and Appendix~\ref{app:h_gate} illustrate the nonadiabatic case.

At a high level, the Z and CZ gates rely on an intentionally engineered avoided crossing between specific Floquet modes that arises due to frequency modulation. By carefully tuning the drive frequency, the system is brought near this avoided crossing in a controlled, adiabatic fashion. As shown by the purple trajectory in Fig.~\ref{fig:general_framework}(d), the system adiabatically follows a Floquet mode to the avoided crossing and back, thereby accumulating a phase determined by both the drive parameters and the crossing structure. In the Z gate, this phase directly implements a rotation about the Z axis. In the CZ gate, the same mechanism is extended to a two-qubit system, where the phase accumulation depends on the control qubit’s state, yielding a conditional phase. In both cases, the gate fidelity hinges on maintaining adiabatic evolution and achieving the desired phase.

In contrast, nonadiabatic gates relax the requirement of adiabaticity, enabling potentially much faster operations. As illustrated by the orange trajectory in Fig.~\ref{fig:general_framework}(d), transitions between Floquet modes are deliberately driven. The trade-off is a stronger demand on the timing precision of the control waveforms to ensure that the intended transitions occur without residual nonadiabatic error. This approach is particularly effective for gates involving population exchange between states, such as the X and Hadamard gates. It is worth emphasizing that the nonadiabatic part of the nonadiabatic gates takes place during the frequency modulation (stages \circnum{2}, \circnum{3}, \circnum{4}), while the amplitude modulation (stages \circnum{1}, \circnum{2}) still takes the adiabatic approach for the purpose of mapping between $\mathcal{H}$ and $\mathcal{K}$.

\subsection{Pulse optimization strategy} \label{sec:pulse_opt_strategy}
\subsubsection{Adiabatic gate} \label{sec:pulse_opt_quasi}
Before further optimization for other parameters, we first choose a set of appropriate control parameters, $\Omega_1,\omega_0$. Let $\omega^*$ be the frequency at which the exact intended avoided crossing takes place for the Floquet mode of interest during the frequency modulation. As a general guideline, these control parameters should be determined so that the quasienergy spectrum during the gate is away from unintended avoided crossings to ensure adiabaticity. The initial drive frequency $\omega_0$ is chosen based on two criteria: 1) During stages \circnum{1}, \circnum{5}, when $\omega_\text{inst}(t) = \omega_0$, the quasienergy spectrum remains distant from avoided crossings as a function of $\Omega(t)$, ensuring adiabaticity during the amplitude modulation; 2) During stages \circnum{2}, \circnum{4}, when $\omega_\text{inst}(t)$ goes from $\omega_0$ to $\omega^*$, the spectrum should remain distant from any other avoided crossings near the path, except for the intended one at $\omega^*$. The drive amplitude $\Omega_1$ is chosen to trade off two competing considerations. On the one hand, increasing $\Omega_1$ requires a longer time $t_a$ to maintain the same level of adiabaticity during amplitude modulation. On the other hand, decreasing $\Omega_1$ reduces the strength of the AC Stark shift (for the Z gate) or ZZ interaction (for the CZ gate) during the frequency modulation, leading to a longer time $2t_\omega+t_h$ required to accumulate the desired phase. Depending on application requirements and other system constraints, an intermediate value of $\Omega_1$ may offer a suitable balance. In the always-on CZ gate in Section~\ref{sec:always_on_CZ}, $\Omega_1$ (along with $\omega_0$) is determined such that the idling ZZ interaction is zero. 

We proceed to design both the amplitude- and frequency-modulation profiles, characterized by the control parameter $\omega_1$ and time parameters $t_a$, $t_\omega$, and $t_h$. To construct the time-dependent drive profiles $\Omega(t)$ and $\omega_\text{inst}(t)$ in stages \circnum{1}, \circnum{5} and \circnum{2}, \circnum{4}, we employ the fast quasiadiabatic (FAQUAD) protocol~\cite{martinez-garaot_Fast_2015,wu_Adiabaticity_2020,garcia-ripoll_Quantum_2020a,guery-odelin_Shortcuts_2019}; see Appendix~\ref{app:faquad} for details. In FAQUAD, the rate of change of the control parameter is adapted dynamically: it slows down when the gap between eigenstates is small (such as near avoided crossings) and speeds up when the gap is large. This produces a trajectory that maintains uniform adiabaticity and reduces the total gate time. Specifically, we identify a subspace of instantaneous eigenstates of the Floquet Hamiltonian $\widetilde{H}_F$ (otherwise referred to as Floquet modes), denoted $\{\ket{\widetilde{u}_{m,\alpha}(t)}\}$, which are most relevant to the intended gate operation. We then apply the multi-level FAQUAD protocol to this subspace. The time parameters $t_a, t_\omega$ are chosen to facilitate local maxima of population overlap of the intended Floquet modes. One could also optimize $t_a, t_\omega$ to meet a target adiabaticity threshold during their respective stages. The time parameter $t_h$ is determined by minimizing the control error $1-\mathcal{F}$ of the intended gate seeking to only constrain unwanted level transitions. Throughout this procedure, the control parameter $\omega_1$ is treated as a hyperparameter to be iterated over. In the end, a specific value of $\omega_1$ is selected to yield a high-fidelity gate taking into account the desired phase.

\subsubsection{Nonadiabatic gate}
The nonadiabatic gate follows the same five-stage structure as the adiabatic gate, but both the control principles and the optimization strategy during the frequency modulation are fundamentally different. We begin by selecting a set of control parameters: $\Omega_1$, $\omega_0$. Since stages \circnum{1}, \circnum{5} still employ adiabatic amplitude modulation, the choices of $\Omega_1$ and $\omega_0$ are guided by similar considerations as in the adiabatic case described in Section~\ref{sec:pulse_opt_quasi}, with necessary modifications for the nonadiabatic regime. Besides considering the adiabaticity and interaction strength, $\Omega_1$ (with $t_a$) also serves as an outer-loop control knob to tune the gate fidelity landscape.

We now turn to the design of the amplitude- and frequency-modulation profiles. The time-dependent drive amplitude $\Omega(t)$ during stages \circnum{1}, \circnum{5} is constructed using the multi-level FAQUAD protocol. During stages \circnum{2}, \circnum{3}, \circnum{4}, three key parameters are $\omega_1$, $t_\omega$ and $t_h$. The objective is to engineer the evolution such that the state in the relevant subspace of $\mathcal{K}$ transitions from ${\ket{\widetilde{u}_{m,\alpha}(t_a)}}$ to $\ket{\widetilde{u}_{m',\alpha'}(t_g - t_a)}$ (with some phase) at the end of the frequency modulation. In principle, one could optimize the full profile $\omega_\text{inst}(t)$ directly. To reduce computational cost while preserving gate fidelity, we instead use a parameterized model $\omega_\text{inst}(t) = f(t; \mathbf{x})$ and optimize the ansatz parameters $\mathbf{x}$ along with $\omega_1$, $t_\omega$, and $t_h$.

\subsubsection{Summary}
We summarize the pulse optimization procedures employed in this work for the design of both adiabatic and nonadiabatic gates.

\begin{center}
\fbox{%
  \parbox{1.0\linewidth}{%
    \textbf{Pulse Optimization Procedure (Adiabatic)} \vspace{0.2em}
    \begin{algorithmic}[1]
    \Require Specific desired gate (and desired phase $\phi^*$ if applicable)  (e.g., phase gate with $\phi^*=\pi$)
    \Require Pre-determined control parameters $\Omega_1, \omega_0$
    \Ensure Optimized control parameters $\{\omega_1, t_a, t_\omega, t_h\}$ and time-dependent drive profiles $\Omega(t)$ and $\omega_\text{inst}(t)$
    \State Construct subspace $\{\ket{\widetilde{u}_{m,\alpha}(t)}\}$ of $\widetilde{H}_F$ relevant to the intended gate
    \State Apply multi-level FAQUAD to design $\Omega(t)$ and optimize $t_a$ for local maxima of adiabaticity
    \For{each candidate $\omega_1$ along the frequency path from $\omega_0$ to $\omega^*$}
        \State \hangindent=2em Apply multi-level FAQUAD to design$\omega_\text{inst}(t)$ and  optimize $t_\omega$ for local maxima of adiabaticity
        \State Optimize $t_h$ to minimize control error $1-\mathcal{F}$ without considering the desired phase $\phi^*$
        \State Compute the final control error $1-\mathcal{F}$ considering the desired phase $\phi^*$
    \EndFor
    \State \hangindent=0em Find the optimized $\omega_1$ to minimize the final control error $1-\mathcal{F}$ considering the desired phase $\phi^*$
    \State \Return Optimized $\{\omega_1, t_a, t_\omega, t_h\}$, $\Omega(t)$ and $\omega_\text{inst}(t)$
    \end{algorithmic}
  }%
}
\end{center}

\begin{center}
\fbox{%
  \parbox{1.0\linewidth}{%
    \textbf{Pulse Optimization Procedure (Nonadiabatic)} \vspace{0.2em}
    \begin{algorithmic}[1]
    \Require Specific desired gate  (e.g., X, Hadamard)
    \Require Pre-determined control parameters $\Omega_1, \omega_0$
    \Require Ansatz parameters $\bf{x}$ for $\omega_\text{inst}(t)= f(t; \mathbf{x})$
    \Ensure Optimized control parameters $\{\omega_1, t_a, t_\omega, t_h\}$ and time-dependent drive profiles $\Omega(t)$ and $\omega_\text{inst}(t)$
    \State Construct subspace $\{\ket{\widetilde{u}_{m,\alpha}(t)}\}$ of $\widetilde{H}_F$ relevant to the intended gate
    \State Numerically scan $\{\omega_1, t_\omega, t_h\}$ and $\bf{x}$ for top candidates in terms of desired population transfer among relevant Floquet modes
    \For{a range of $t_a$}
        \State \hangindent=2em Apply multi-level FAQUAD to design $\Omega(t)$ and $t_a$ and compute the final control error $1-\mathcal{F}$
    \EndFor
    \State \hangindent=0em Find the optimized $\Omega(t)$ and $t_a$ to minimize the final control error $1-\mathcal{F}$
    \State \Return Optimized $\{\omega_1, t_a, t_\omega, t_h\}$, $\Omega(t)$ and $\omega_\text{inst}(t)$
    \end{algorithmic}
  }%
}
\end{center}

\section{Adiabatic gates} \label{sec:adiabatic_gates}
In this section, we present three representative examples of adiabatic gates: the CZ gate, its always-on variant, and the single-qubit Z gate, all constructed using the proposed framework. These examples illustrate the flexibility of the framework while highlighting the trade-offs between speed and control errors that characterize adiabatic protocols.

\begin{figure*}[htbp]
    \centering
    \includegraphics[width=1.01\linewidth]{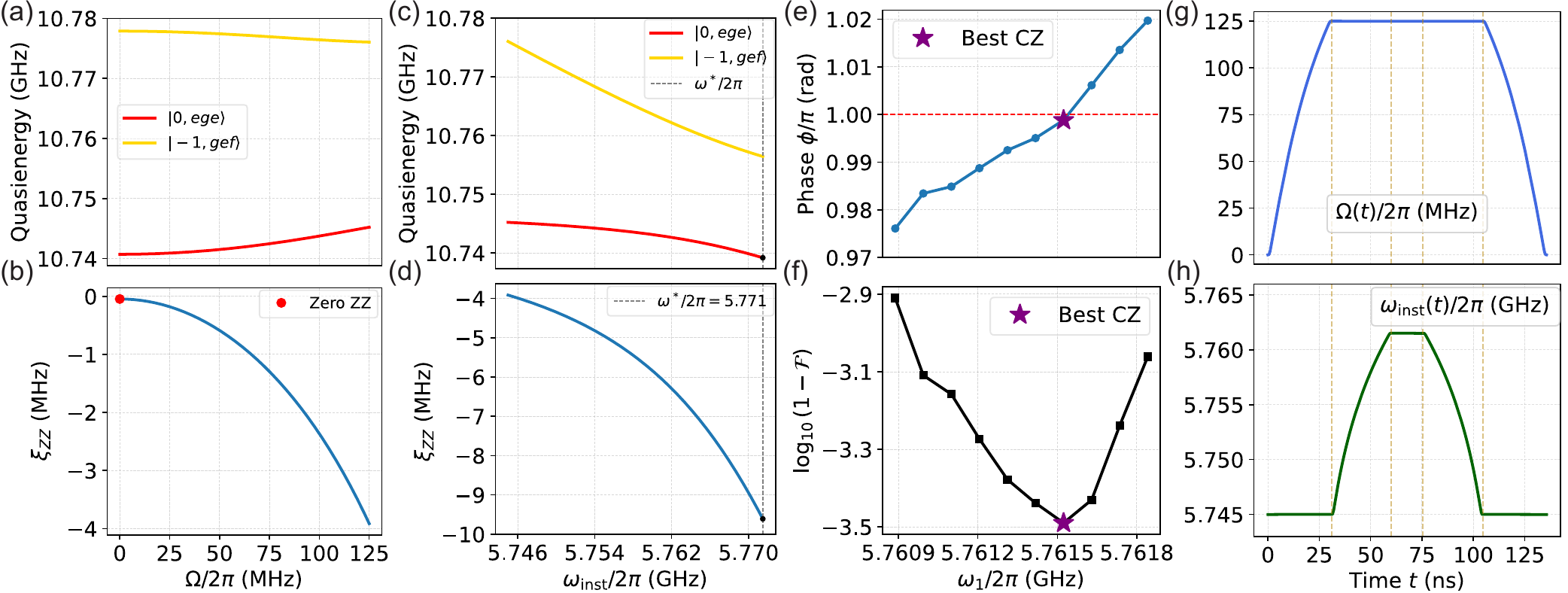}
    \caption{ CZ gate. (a)(c) Quasienergy spectrum of the computational Floquet mode $\ket{0,ege}$ and non-computational Floquet mode $\ket{-1,gef}$ as a function of drive amplitude $\Omega(t)$ and instantaneous frequency $\omega_\mathrm{inst}(t)$. $\omega^*/2\pi=5.771$ GHz indicates where the closest point in quasienergy takes place, which also dictates the maximal allowed ZZ interaction $-9.61$ MHz. (b)(d) ZZ interaction as a function of drive amplitude $\Omega(t)$ and instantaneous frequency $\omega_\text{inst}(t)$. (e)(f) Phase accumulation $\phi$ and control error of CZ gates as a function of instantaneous frequency endpoint $\omega_1$. Purple star indicates the best CZ gate among the samples shown. (g)(h) The optimized control waveforms $\Omega(t)$ and $\omega_\text{inst}(t)$ for the best CZ gate.}
    \label{fig:cz_gate}
\end{figure*}

\subsection{Two-qubit gate} \label{sec:two_qubit_gate}
As a first example, we apply the principles of adiabatic gates to the two-qubit CPHASE gate, with a particular focus on the CZ gate. The CZ gate is represented in matrix form as
\begin{equation} \label{eq:1}
U_{\textrm{CZ}} =
\begin{bmatrix}
1 & 0 & 0 & 0 \\
0 & 1 & 0 & 0 \\
0 & 0 & 1 & 0 \\
0 & 0 & 0 & -1
\end{bmatrix}.
\end{equation}

We consider a system described by the Hamiltonian
\begin{equation}
\begin{split}
\hat{H} = &\sum_{j} \omega_j \hat{a}_j^\dagger \hat{a}_j + \frac{\alpha_j}{2}\hat{a}_j^\dagger\hat{a}_j^\dagger \hat{a}_j\hat{a}_j \\
&+ \sum_{j\neq k} -g_{jk}(\hat{a}_j - \hat{a}^\dagger_j)(\hat{a}_k - \hat{a}^\dagger_k) \\
&+ \Omega(t)\cos[\theta(t)]\hat{n}_c,
\end{split}
\end{equation}
where the CZ gate is implemented by applying a microwave drive to the coupler. As the drive frequency is modulated, a significant ZZ interaction is activated, leading to the accumulation of a controlled phase. We simulate the CZ gate up to single-qubit rotations.
We propose two methods for implementing such a CZ gate. The first is a standard CZ gate, where the drive is turned on and off at the beginning and end of the gate. The second is an always-on CZ gate, where the drive remains on as part of the idling system, and gate activation is solely achieved through frequency modulation. The latter can be regarded as a special case of the former and offers the advantage of eliminating the time overhead associated with amplitude modulation, resulting in a shorter overall gate duration. We will use the notation $\ket{\text{QBa},\text{CPLRc},\text{QBb}}$ to denote the eigenstates of the system, with $\text{QBa},\text{CPLRc},\text{QBb}\in\{g,e,f,h,\dots\}$. The ZZ interaction is defined as
\begin{equation}
    \xi_{ZZ}=(\varepsilon_{m,ege}+\varepsilon_{m,ggg}-\varepsilon_{m,gge}-\varepsilon_{m,egg})/2\pi,
\end{equation}
where $\varepsilon_{m,\text{QBa},\text{CPLRc},\text{QBb}}$ is the quasienergy of the Floquet mode $\ket{m,\text{QBa},\text{CPLRc},\text{QBb}}$, expressed in the explicit notation that identifies both the Floquet photon number $m$ and the bare basis state $\ket{\text{QBa},\text{CPLRc},\text{QBb}}$.

\subsubsection{CZ gate} \label{sec:traditional_CZ}
We begin by demonstrating the CZ gate in the qubit-coupler-qubit system. The device parameters used in this example are summarized in Table~\ref{tab:CZ_gate_params} in Appendix~\ref{app:conventional_CZ}. We design the drive amplitude and initial drive frequency to be $\Omega_1/2\pi=125$ MHz and $\omega_0/2\pi=5.745$ GHz, following guidance given in Section~\ref{sec:pulse_opt_quasi}. As we ramp up the drive amplitude $\Omega(t)$, the quasienergy spectrum of the computational Floquet mode $\ket{0,ege}$ and its closest non-computational Floquet mode $\ket{-1,gef}$ is shown in Fig.~\ref{fig:cz_gate}(a). As depicted in Fig.~\ref{fig:cz_gate}(b), the ZZ interaction vanishes at $\Omega=0$ (red dot) for this set of device parameters and gradually departs from zero with increasing $\Omega(t)$. Figures~\ref{fig:cz_gate}(c)--(d) show the quasienergy spectrum and the ZZ interaction as a function of $\omega_\text{inst}(t)$. As $\ket{0,ege}$ and $\ket{-1,gfe}$ get closer in quasienergy during the frequency modulation, the ZZ interaction becomes larger. The frequency $\omega^*/2\pi=5.771$ GHz is where the exact avoided crossing takes place, resulting in a maximum ZZ interaction of about $-9.61$ MHz since we require that $\omega_\text{inst}(t)\leq\omega^*$. We design both $\Omega(t)$ and $\omega_\text{inst}(t)$ following the protocols outlined in Section~\ref{sec:pulse_opt_quasi}. Figures~\ref{fig:cz_gate}(e)--(f) show the CZ gate phase accumulation $\phi$ by the state $\ket{eeg}$ and control error $1-\mathcal{F}$ as a function of samples of $\omega_1$. The best CZ gate of the samples shown is indicated by the purple star where $\omega_1/2\pi=5.7615$ GHz, whose CZ gate control error approaches $0.03\%$ and total gate duration is $135.6$ ns. The optimized $\Omega(t)$ and $\omega_\text{inst}(t)$ are shown in Figs.~\ref{fig:cz_gate}(g)--(h).
We note that the phase accumulation $\phi$ in Fig.~\ref{fig:cz_gate}(e) is expected to be a continuum, and therefore there exists some  $\omega_1$ that exactly hits $\phi=\pi$. The same argument is true for the other gates shown in Sections~\ref{sec:always_on_CZ} and~\ref{sec:Z_gate}. More details can be found in Appendix~\ref{app:conventional_CZ}. 

\begin{figure}[htbp]
    \centering
    \includegraphics[width=1.01\linewidth]{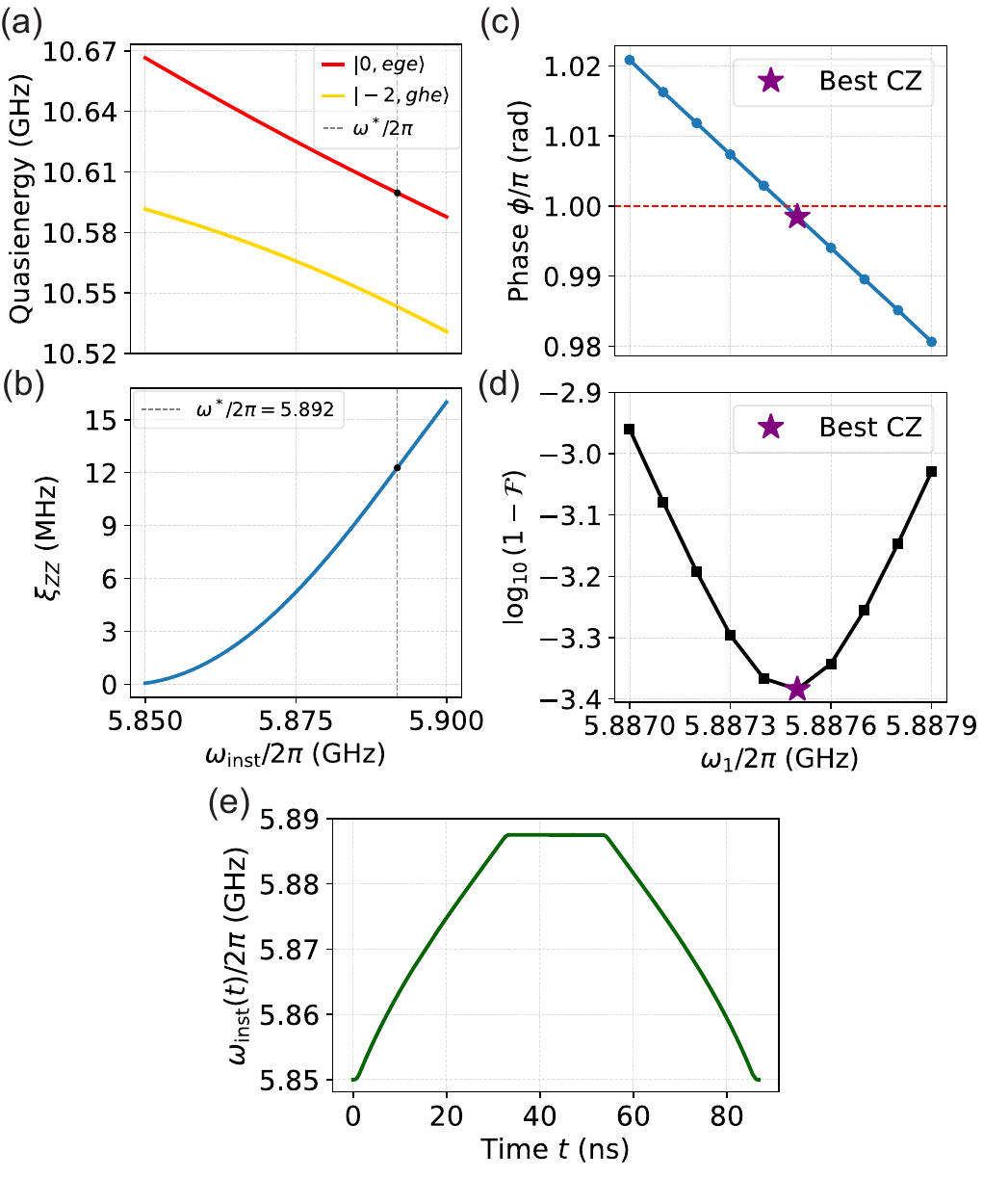}
    \caption{Always-on CZ gate. (a) Quasienergy spectrum of the computational Floquet mode $\ket{0,ege}$ coming into avoided crossing with non-computational Floquet mode $\ket{-2,ghe}$ as a function of instantaneous frequency $\omega_\mathrm{inst}(t)$. $\omega^*/2\pi=5.892$ GHz indicates where the closest point in quasienergy takes place, which also dictates the maximal allowed ZZ interaction $12.3$ MHz. (b) ZZ interaction as a function of instantaneous frequency $\omega_\mathrm{inst}(t)$. (c)(d) Phase accumulation and control error of always-on CZ gates as a function of instantaneous frequency endpoint $\omega_1$. Purple star indicates the best CZ gate among the samples shown. (e) The optimized control waveform $\omega_\text{inst}(t)$ for the best CZ gate.}
    \label{fig:alwayson_cz_gate}
\end{figure}

\subsubsection{Always-on CZ gate} \label{sec:always_on_CZ}
Compared to the CZ gate presented above, a key feature of the always-on CZ gate is that the drive remains an essential component of the system even during idle periods. The control of driven qubits has been the subject of extensive study~\cite{wolk_Quantum_2017, timoney_Quantum_2011, tan_Demonstration_2013, li_Motional_2013, bermudez_Robust_2012, mundada_FloquetEngineered_2020, huang_Engineering_2021a, mikelsons_Universal_2015, gandon_Engineering_2022}. The computational basis states in driven qubits are defined to be the Floquet modes of the driven system. Each Floquet mode can be mapped to the bare eigenstates of the undriven system by adiabatically ramping down the drive amplitude.

We present an example of the always-on CZ gate using the device parameters summarized in Table~\ref{tab:alwayson_CZ_gate_params} in Appendix~\ref{app:alwayson_CZ}. Under this set of device parameters, the static ZZ interaction (without drive) is $-0.368$ MHz, which which leads to unwanted phase accumulations if not properly addressed. Therefore, we design the always-on drive amplitude and frequency to be $\Omega_1/2\pi=254.1$ MHz and $\omega_0/2\pi=5.85$ GHz so that the idling ZZ interaction (with drive) vanishes. Note that these two parameter choices are not unique, but the final gate fidelity may depend on the parameter choices. In the always-on CZ gate, there is no explicit amplitude modulation, i.e., stages \circnum{1}, \circnum{5} do not exist. As shown in Figs.~\ref{fig:alwayson_cz_gate}(a)--(b), the Floquet modes $\ket{0,ege}$ and $\ket{-2,ghe}$ come into an avoided crossing as $\omega_\text{inst}(t)$ is modulated, during which the ZZ interaction increases. The dashed vertical line indicates the exact avoided crossing at $\omega^*/2\pi=5.982$ GHz, where we have maximum ZZ interaction of about $12.3$ MHz since we limit $\omega_\text{inst}(t)\leq\omega^*$. We note that in the single excitation manifold, the Floquet modes $\ket{0,gge},\ket{0,egg}$ are also relatively close in quasienergies to $\ket{-2,gfe},\ket{-2,ghg}$, which we include in the multi-level FAQUAD design for $\omega_1,t_\omega,t_h$ and $\omega_\mathrm{inst}(t)$. Figures~\ref{fig:alwayson_cz_gate}(c)--(d) show the CZ gate phase accumulation $\phi$ by the Floquet mode $\ket{0,eeg}$ and control error $1-\mathcal{F}$ as a function of samples of $\omega_1$. The best CZ gate of the shown samples is the one with $\omega_1/2\pi=5.8875$ GHz (purple star in Figs.~\ref{fig:alwayson_cz_gate}(c)--(d)), which has a CZ gate control error of $0.04\%$ and a total gate duration of $86.8$ ns. The optimized $\omega_\text{inst}(t)$ is shown in Fig.~\ref{fig:alwayson_cz_gate}(e). Additional details can be found in Appendix~\ref{app:alwayson_CZ}. 

\subsection{Single-qubit gate} \label{sec:Z_gate}
\subsubsection{Phase gate (Z, S, T)}
\begin{figure*}[htbp]
    \centering
    \includegraphics[width=1.01\linewidth]{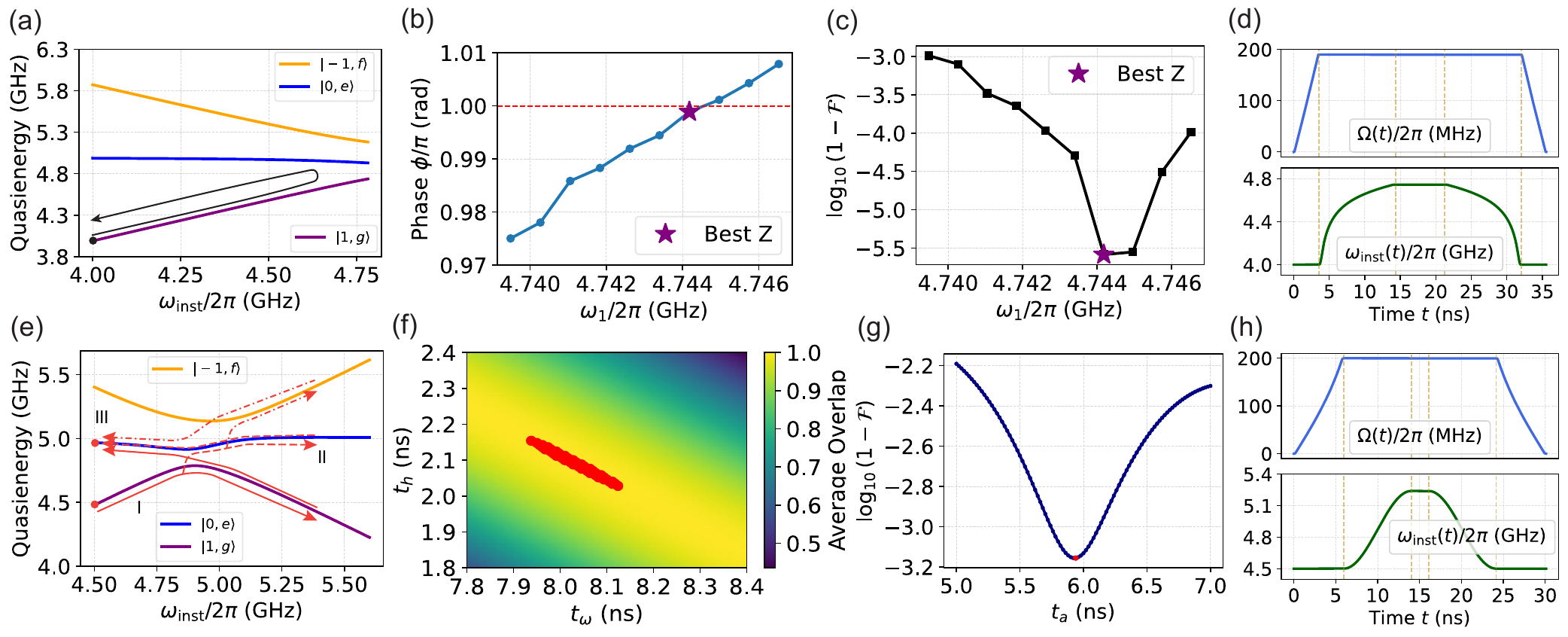}
    \caption{Z gate (a-d) and X gate (e-h). (a) Quasienergy spectrum of the Floquet modes $\{\ket{1,g},\ket{0,e},\ket{-1,f}\}$-subspace as a function of $\omega_\text{inst}(t)$ for the Z gate. The black curve indicates that the population stays in $\ket{1,g}$ during the Z gate if initialized in $\ket{g}$. (b)(c) Phase accumulation and control error of Z gates as a function of instantaneous frequency endpoint $\omega_1$. Purple star indicates the best Z gate among the samples shown. (d) The optimized control waveforms $\Omega(t)$ and $\omega_\text{inst}(t)$ for the best Z gate. (e) Quasienergy spectrum of the Floquet modes $\{\ket{1,g},\ket{0,e},\ket{-1,f}\}$-subspace as a function of $\omega_\text{inst}(t)$ for the X gate. The red curves (solid, dashed, dash-dot line) show the population transitions during the frequency modulation of the X gate if initialized in $\ket{g}$. Part I: the population remains in $\ket{1,g}$. Part II: the population transfers to $\ket{1,g}$, $\ket{0,e}$ and $\ket{-1,f}$. Part III: the population goes to $\ket{0,e}$. (f) Average overlap as a function of $t_\omega,t_h$ when $\omega_1/2\pi=5.24$ GHz at the end of the frequency modulation during the X gate. Red area in the middle indicates top candidates for potential high-fidelity X gates. (g) The X gate control error as a function of $t_a$ for one specific set of $t_\omega,t_h,\omega_1$ among the top candidates in (f). (h) The optimized control waveforms $\Omega(t)$ and $\omega_\text{inst}(t)$ for the best X gate. }
    \label{fig:x_z_gate}
\end{figure*}
Single-qubit phase gates can be implemented in a similar manner using adiabatic principles. The phase gate imparts a relative phase \( \phi \) between the qubit \(|g\rangle\) and \(|e\rangle\) states, and is represented as
\begin{equation}
    \mathrm{R}_z(\phi) = \begin{bmatrix} e^{-i\phi/2} & 0 \\ 0 & e^{i\phi/2} \end{bmatrix}.
\end{equation}
In particular, we focus on the Z gate, which corresponds to the special case \( \phi^* = \pi \). We numerically simulate the Z gate up to a global phase. It is worth noting that the S and T gates are other special cases of the phase gate, and the gate principles are the same as the Z gate, but the durations of the S and T gates are shorter.

We consider a system of a driven transmon qubit with Hamiltonian 
\begin{equation} \label{eq:1qb_H}
    \hat{H} = \omega_q \hat{a}^\dagger \hat{a} + \frac{\alpha}{2}\hat{a}^\dagger\hat{a}^\dagger\hat{a}\hat{a} + \Omega(t)\cos[\theta(t)]\hat{n}.
\end{equation}
The corresponding Floquet Hamiltonian $\widetilde{H}_F(t)$ in the extended Hilbert space $\mathcal{K}$ can be found as in Eq.~\eqref{eq:floquet_H_compact}.
To illustrate the working principles of the Z gate, we consider the lowest three energy levels for the transmon qubit $\ket{g}, \ket{e}, \ket{f}$ and specifically analyze the $\{\ket{1,g}, \ket{0,e},\ket{-1,f}\}$-subspace of $\widetilde{H}_F(t)$ with Hamiltonian
\begin{equation} \label{eq:1qb_K_sub}
    \displaystyle \widetilde{H}_{F,sub}(t) = \begin{bmatrix} \omega_{\text{inst}}(t) & \displaystyle\frac{\Omega(t)}{2} & 0 \\ \displaystyle\frac{\Omega(t)}{2}  & \omega_q & \displaystyle\frac{\sqrt{2}\Omega(t)}{2} \\ 0 & \displaystyle\frac{\sqrt{2}\Omega(t)}{2} & 2\omega_q+\alpha -\omega_{\text{inst}}(t) \end{bmatrix}.
\end{equation}
The reasoning behind this is that all the other Floquet modes $\{\ket{m+1,g}, \ket{m,e},\ket{m-1,f}\}, m\neq0$ can be considered as replicas of this subspace due to the degeneracies of different Floquet-Brillouin zones (see Appendix~\ref{app:basic_floquet}).

As shown by the black curve in Fig.~\ref{fig:x_z_gate}(a), the Z gate is implemented in a way where $\omega_\text{inst}(t)$ is varied close to (but not across) the avoided crossing region and just enough to accumulate a desired phase. We present an example using the device and (partial) control parameters summarized in Table~\ref{tab:Z_gate_params} in Appendix~\ref{app:z_gate}. We proceed to design the remaining parameters—$\omega_1$, $t_a$, $t_\omega$, $t_h$—and the time-dependent drive profiles $\Omega(t)$ and $\omega_\text{inst}(t)$, following the adiabatic pulse optimization procedure outlined in Section~\ref{sec:pulse_opt_quasi}. Figures~\ref{fig:x_z_gate}(b)--(c) show the phase accumulation $\phi$ by the state $\ket{e}$ and the Z gate control error $1-\mathcal{F}$ as a function of samples of $\omega_1$. The best Z gate of the shown samples is the one with $\omega_1/2\pi=4.7442$ GHz (purple star in Figs.~\ref{fig:x_z_gate}(b)--(c)), which has a Z-gate control error below $3 \times 10^{-6}$ and total gate duration of $35.5$ ns. The optimized control waveforms $\Omega(t)$ and $\omega_\text{inst}(t)$ are shown in Fig.~\ref{fig:x_z_gate}(d). Additional details can be found in Appendix~\ref{app:z_gate}.

\section{Nonadiabatic gates} \label{sec:nonadiabatic_gates}
In this section, we present nonadiabatic examples constructed using the proposed framework to complete a universal gate set based primarily on frequency modulation. Although the resulting protocols are more intricate than conventional Rabi-drive gates, they demonstrate the possibility of harnessing nonadiabatic transitions within the Floquet framework and may find applications where fast, nonadiabatic control offers advantages.

\subsection{X gate}\label{sec:XY_gate}
As a representative example of the nonadiabatic gate, we simulate the X gate, whose resulting operation is represented by the matrix:
\begin{equation}
    \mathrm{X}_{\phi_1,\phi_2} = \begin{bmatrix} 0 & e^{i\phi_1} \\ e^{i\phi_2} & 0 \end{bmatrix}.
\end{equation}
Similar techniques can be used to implement any X/Y gate with arbitrary rotational phase.

% We consider a system of a driven transmon qubit with Hamiltonian 
% \begin{equation} \label{eq:1qb_H}
%     \hat{H} = \omega_q \hat{a}^\dagger \hat{a} + \frac{\alpha}{2}\hat{a}^\dagger\hat{a}^\dagger\hat{a}\hat{a} + \Omega(t)\cos[\theta(t)]\hat{n}.
% \end{equation}
% The corresponding Floquet Hamiltonian $\widetilde{H}_F(t)$ in the extended Hilbert space $\mathcal{K}$ can be found as in Eq.~\eqref{eq:floquet_H_compact}.
% To illustrate the working principles of the X gate, we consider the bottom three energy levels for the transmon qubit $\ket{g}, \ket{e}, \ket{f}$ and further the $\{\ket{1,g}, \ket{0,e},\ket{-1,f}\}$-subspace of $\widetilde{H}_F(t)$ with Hamiltonian
% \begin{equation} \label{eq:1qb_K_sub}
%     \displaystyle \widetilde{H}_{F,sub}(t) = \begin{bmatrix} \omega_{\text{inst}}(t) & \displaystyle\frac{\Omega(t)}{2} & 0 \\ \displaystyle\frac{\Omega(t)}{2}  & \omega_q & \displaystyle\frac{\sqrt{2}\Omega(t)}{2} \\ 0 & \displaystyle\frac{\sqrt{2}\Omega(t)}{2} & 2\omega_q+\alpha -\omega_{\text{inst}}(t) \end{bmatrix}.
% \end{equation}
% The reasoning behind this is that all the other Floquet modes $\{\ket{m+1,g}, \ket{m,e},\ket{m-1,f}\}, m\neq0$ can be considered as replicas of this subspace due to the degeneracies of different Floquet-Brillouin zones (see Appendix~\ref{app:basic_floquet}).

The system Hamiltonian $\hat{H}$ and Floquet Hamiltonian $\widetilde{H}_F(t)$ are the same as in Section~\ref{sec:Z_gate}. 
Figure~\ref{fig:x_z_gate}(e) shows the quasienergy spectrum of the Floquet $\{\ket{1,g}, \ket{0,e},\ket{-1,f}\}$-subspace as a function of $\omega_\text{inst}(t)$. The red curves indicate the evolution of the qubit population if it was initialized in $\ket{g}$. The goal of the frequency modulation during the X gate is to drive a complete population transfer between $\ket{1,g}$ and $\ket{0,e}$ while minimizing leakage to $\ket{-1,f}$ at the end of the evolution. We use the device and (partial) control parameters summarized in Table~\ref{tab:X_gate_params} in Appendix~\ref{app:x_y_gate}. In particular, for a given set of device parameters $(\omega_q, \alpha)$, there exist multiple viable choices for control parameters $\Omega_1$ and $(\omega_0, \omega_1)$ that, together with the remaining pulse parameters, yield a fast and high-fidelity X gate.

The time parameter $t_a$ and the amplitude modulation profile $\Omega(t)$ are designed using the multi-level FAQUAD protocol~\cite{martinez-garaot_Fast_2015,wu_Adiabaticity_2020,garcia-ripoll_Quantum_2020a,guery-odelin_Shortcuts_2019}. For the design of the frequency modulation profile $\omega_\text{inst}(t)$, we use a simple model parameterized by $\omega_1,t_\omega,t_h$. The instantaneous frequency $\omega_\text{inst}(t)$ follows a piecewise-defined symmetric trajectory: it goes from $\omega_0$ to $\omega_1$ over a duration $t_\omega$ using a half-cosine profile, remains at $\omega_1$ for a hold time $t_h$, and finally goes from $\omega_1$ to $\omega_0$ over a duration $t_\omega$ using another half-cosine profile. Note that during the frequency modulation there can be population leakage to $\ket{-1,f}$, but eventually the population will mostly be in the computational subspace. We define an average overlap quantity as the average of the final population overlap in $\ket{0,e}$ if initialized in $\ket{1,g}$ and the final population overlap in $\ket{1,g}$ if initialized in $\ket{0,e}$. We scan $t_\omega,t_h,\omega_1$ for the best average overlap. Figure~\ref{fig:x_z_gate}(f) shows the average overlap as a function of $t_\omega,t_h$ when $\omega_1/2\pi=5.24$ GHz near the optimal region. We then pick the top candidates within the region (indicated by the red area) and simulate the X gate as a function of different amplitude ramp-up times $t_a$ and the corresponding amplitude profile $\Omega(t)$. The X gate control errors for a specific set of $t_\omega,t_h,\omega_1$ are shown in Fig.~\ref{fig:x_z_gate}(g), where the red dot indicates an optimized value of $t_a$ that gives the best X gate. We find that this best X gate has a control error of $0.07\%$ and a total gate duration of $30.1$ ns. The optimized control waveforms $\Omega(t)$ and $\omega_\text{inst}(t)$ are shown in Fig.~\ref{fig:x_z_gate}(h). Additional details can be found in Appendix~\ref{app:x_y_gate}.

\subsection{Hadamard gate}
An example of the Hadamard gate, obtained using the same procedure as for the X gate, is presented in Appendix~\ref{app:h_gate}. The best Hadamard gate we find has a control error of approximately $0.07\%$ and a total gate duration of $37.4$ ns.

\section{Conclusion and outlook} \label{sec:conclusion}
In this work, we develop a general theoretical framework for high-fidelity quantum gate design based on frequency- and amplitude-modulated drives. We extend conventional amplitude modulation by incorporating frequency modulation as an additional control dimension. In addition, our framework accommodates the design of both adiabatic and nonadiabatic gates. Numerical simulations with realistic transmon parameters demonstrate that this framework supports a universal gate set—including the X, Hadamard, phase, and CZ gates—with control errors below 0.1\%. The resulting gate durations are 25--40 ns for single-qubit X, Hadamard and Z gates, and 80--135 ns for two-qubit CZ gate and its always-on version. The feasibility of hardware implementations of these control schemes is discussed in Appendix~\ref{app:hardware}.

Beyond the specific examples of different gate operations presented, the proposed framework expands the toolbox of microwave-based quantum control and provides a systematic optimization strategy for parameters of gate design. In principle, this framework supports fully general gate protocols that trace arbitrary trajectories in the $\omega_\text{inst}(t)-\Omega(t)$ parameter space. In our work, amplitude and frequency modulation are applied in separate stages, providing a structured approach that is simpler to understand and implement while still capturing the essential features of the broader framework. While our analysis focuses on transmon qubits modeled as Kerr nonlinear oscillators, the proposed framework can be readily extended to capture the full transmon nonlinearity and is potentially applicable to other qubit modalities, such as fluxonium qubits, neutral atoms, or trapped ions, where frequency-modulated control is feasible.

Looking ahead, the framework can be extended to develop optimized control protocols for larger qubit systems, since the underlying extended Hilbert space representation naturally accommodates multiple qubits and couplers with multiple control channels, as briefly discussed in Appendix~\ref{app:generalized_Floquet}. Incorporating robustness against noise and parameter variations is another important direction. We anticipate that frequency- and amplitude-modulated control will provide valuable perspectives for quantum gate design and control in the pursuit of scalable, high-fidelity quantum processors.

\section*{Acknowledgments}
We gratefully acknowledge insightful conversations with Thomas A. Baran, Petros T. Boufounos, Chungwei Lin, Terry P. Orlando, Kyle Serniak and Helin Zhang. This research was funded in part by the Army Research Office under Award Number W911NF-23-1-0045 and in part by the U.S. Department of Energy, Office of Science, National Quantum Information Science Research Centers, Quantum Systems Accelerator. Additionally, S.C. acknowledges support from the NSF Graduate Research Fellowship. The views and conclusions contained herein are those of the authors and should not be interpreted as necessarily representing the official policies or endorsements, either expressed or implied, of the U.S. Government.

%\newpage
\appendix
\section{Floquet theory} \label{app:floquet}
Floquet theory is a mathematical framework used to analyze differential equations with periodic coefficients~\cite{floquet_Equations_1883}, and is especially common in quantum systems under time-periodic driving.
\subsection{Basic idea} \label{app:basic_floquet}
We shall be interested in quantum systems with Hamiltonians that are periodic in time: $\hat{H}(t+T)=\hat{H}(t)$. Here, $T$ is the time period. We can equivalently define the drive frequency $\omega=2\pi/T$. The system is described by the Schr\"odinger equation
\begin{equation} \label{eq:sch_eq}
     i \frac{\partial}{\partial t}\ket{\psi(t)}=\hat{H}(t) \ket{\psi(t)}.
\end{equation}
According to Floquet theory, there exist solutions to Eq.~\eqref{eq:sch_eq} of the form
\begin{equation} \label{eq:sol_sch_eq}
    \ket{\psi_\alpha(t)} = e^{-i\varepsilon_\alpha t}\ket{u_\alpha(t)},
\end{equation}
where $\ket{u_\alpha(t)}$ is periodic in time, i.e., $\ket{u_\alpha(t+T)}=\ket{u_\alpha(t)}$. Here, $\varepsilon_\alpha$ is real-valued and referred to as the quasienergy. $\ket{u_\alpha(t)}$ is referred to as the Floquet mode, while $\ket{\psi_\alpha(t)}$ is referred to as the Floquet state. Here, $\alpha$ is an index denoting the qubit state ranging from $0$ to $N-1$, where $N$ is the dimension of the Hilbert space considered. The Floquet modes $\{\ket{u_\alpha(t)}\}$ form a complete basis and therefore the general solution to Eq.~\eqref{eq:sch_eq} can be written as a linear combination of the form:
\begin{equation}
    \ket{\psi(t)} = \sum_\alpha c_\alpha \ket{\psi_\alpha(t)}=\sum_\alpha c_\alpha e^{-i\varepsilon_\alpha t}\ket{u_\alpha(t)}.
\end{equation}
We note that for each index $\alpha$, there exists a class of degenerate Floquet modes
\begin{equation}
    \ket{u_{m,\alpha}(t)} = \ket{u_\alpha(t)}e^{im\omega t},
\end{equation}
indexed by $(m,\alpha)$ for $m\in \mathbb{Z}$, which yield an identical solution for the Floquet state $\ket{\psi_\alpha(t)}$ in Eq.~\eqref{eq:sol_sch_eq}, but with shifted quasienergies
\begin{equation}
    \varepsilon_{m,\alpha} = \varepsilon_\alpha +m\omega.
\end{equation}
Thus, all non-degenerate Floquet modes can be found by restricting the corresponding quasienergies in a ``Floquet-Brillouin zone'', where $\varepsilon_{\text{FB}}\leq\varepsilon_\alpha <\varepsilon_{\text{FB}}+\omega$.
\vspace{1em}

\subsection{Extended Hilbert space representation} \label{sec:extended_HS}
In this section, we briefly summarize two approaches to constructing the extended Hilbert space representation~\cite{sambe_Steady_1973,shirley_Solution_1965,eckardt_Highfrequency_2015,rudner_Floquet_2020,silveri_Quantum_2017a,howland_Stationary_1974,chu1985advances}. The first adopts a semiclassical viewpoint, in which the qubit is treated quantum mechanically while the drive field is treated classically. Here, the extended Hilbert space representation serves as a purely mathematical tool, enabling the transformation of the Schr\"odinger equation with a time-dependent Hamiltonian into an equivalent equation with a time-independent Hamiltonian. The second approach is motivated by physical considerations. We also note a comprehensive construction in which both the qubit and the drive field are treated as quantized can be found in~\cite{bialynicki-birula_Quantum_1976,guerin_Complete_1997,guerin_Control_2003}.

We show the first approach. If we substitute Eq.~\eqref{eq:sol_sch_eq} into Eq.~\eqref{eq:sch_eq}, we get 
\begin{equation}\label{eq:flq_eig}
    \bigg[\hat{H}(t) - i \frac{\partial}{\partial t}\bigg]\ket{u_\alpha(t)} = \varepsilon_\alpha \ket{u_\alpha(t)},
\end{equation}
and the task is to find an algebraic way to solve this equation.

Using a Fourier series, we can write $\hat{H}(t) = \displaystyle\sum_{k=-\infty}^{+\infty} \hat{H}^{(k)} e^{ik\omega t}$ and $\ket{u_\alpha(t)} = \displaystyle\sum_{l=-\infty}^{+\infty} e^{-il\omega t} \ket{u_\alpha^{(l)}}$. 

Putting these two equations into Eq.~\eqref{eq:flq_eig}, we will get
\begin{equation}
    \sum_{l=-\infty}^{+\infty} \hat{H}^{(l-n')} \ket{u_\alpha^{(l)}} -n'\omega  \ket{u_\alpha^{(n')}} = \varepsilon_\alpha \ket{u_\alpha^{(n')}},
\end{equation}
for $n'\in \mathbb{Z}$. 

Grouping these equations into a matrix form, we will have the following eigenproblem in the extended Hilbert space
\begin{equation} \label{eq:floquet_eig}
    \widetilde{H}_F \ket{\widetilde u_\alpha} = \varepsilon_\alpha \ket{\widetilde u_\alpha},
\end{equation}
where
\begin{widetext}
\begin{equation} \label{eq:H_F}
    \widetilde{H}_F = \begin{bmatrix}
        \ddots & \vdots & \vdots & \vdots & \vdots & \vdots &  \reflectbox{$\ddots$} \\
        \cdots & \hat{H}^{(0)}+2\omega & \hat{H}^{(1)} & \hat{H}^{(2)} & \hat{H}^{(3)} & \hat{H}^{(4)} & \cdots \\
        \cdots & \hat{H}^{(-1)} & \hat{H}^{(0)}+\omega & \hat{H}^{(1)} & \hat{H}^{(2)} & \hat{H}^{(3)} & \cdots \\
        \cdots & \hat{H}^{(-2)} & \hat{H}^{(-1)} & \hat{H}^{(0)} & \hat{H}^{(1)} & \hat{H}^{(2)} & \cdots \\
        \cdots & \hat{H}^{(-3)} & \hat{H}^{(-2)} & \hat{H}^{(-1)} & \hat{H}^{(0)}-\omega & \hat{H}^{(1)} & \cdots \\
        \cdots & \hat{H}^{(-4)} & \hat{H}^{(-3)} & \hat{H}^{(-2)} & \hat{H}^{(-1)} & \hat{H}^{(0)}-2\omega & \cdots \\ \reflectbox{$\ddots$} 
         & \vdots & \vdots & \vdots & \vdots & \vdots & \ddots
        \end{bmatrix},\ \ket{\widetilde u_\alpha} = \begin{bmatrix} \vdots \\ \ket{u_\alpha^{(-2)}} \\ \ket{u_\alpha^{(-1)}} \\ \ket{u_\alpha^{(0)}} \\ \ket{u_\alpha^{(1)}} \\ \ket{u_\alpha^{(2)}} \\ \vdots
    \end{bmatrix}.
\end{equation}
\end{widetext}

Here, we use $\ket{\widetilde{\ \cdot \ }}$ to denote state vectors in the extended Hilbert space and $\,\widetilde{\cdot}\,$ to denote operators operating on the extended Hilbert space. Once we diagonalize $\widetilde{H}_F$ for $\varepsilon_\alpha$ and $\ket{\widetilde u_\alpha}$, we can reconstruct Floquet modes $\ket{u_\alpha(t)}$ and solve for evolution in the original Hilbert space. It is worth mentioning that the Floquet Hamiltonian $\widetilde{H}_F$ in the extended Hilbert space is of infinite dimension. When performing numerical calculations, $\widetilde{H}_F$ needs to be truncated to some finite dimension satisfying convergence checks.

Next, we adopt a more physics-motivated perspective. As an example, we consider a charge-driven qubit system $\hat{H}(t) = \hat{H}_0+\hat{H}_\text{drive}(t)$, where $\hat{H}_\text{drive}(t)=\Omega\cos{(\omega t)}\hat{n}$ with $\Omega$ and $\omega$ the drive amplitude and frequency, respectively. Let $\theta(t) = \omega t$ and promote $\theta(t)$ to a $2\pi$-periodic quantum degree of freedom $\hat{\vartheta}$ whose conjugate variable is $\hat{m}\rightarrow -i\partial_{\vartheta}$, satisfying $[\hat{\vartheta},\hat{m}]=i$. Therefore, we can write $\hat{H}(t) - i\partial/\partial t$ in Eq.~\eqref{eq:flq_eig} as
\begin{equation}\label{eq:vartheta_map}
    \widetilde{H}_F = \hat{H}(\hat{\vartheta}) + \omega\hat{m} = \hat{H}_0 + \Omega\cos{(\hat{\vartheta})}\hat{n} +\omega\hat{m} ,
\end{equation}
where we have used $\partial/\partial t = \partial/\partial \vartheta \cdot \partial \vartheta/\partial t = \omega\hat{m}$. 
We then represent Eq.~\eqref{eq:vartheta_map} in the $\hat{m}$-basis. Since $\hat{\vartheta}$ is $2\pi$-periodic, $\hat{m}$ must be quantized to have eigenvalues $m\in\mathbb{Z}$, i.e., $\hat{m}=\sum_m m\ket{m}\!\bra{m}$. We can also write $\cos(\hat\vartheta) = \frac{1}{2}\sum_m \ket{m+1}\!\bra{m}{\rm \ +\  h.c.}$. If we plug these into Eq.~\eqref{eq:vartheta_map}, we will explicitly recover Eq.~\eqref{eq:H_F}. An advantage of this perspective is that there is some similarity in Eq.~\eqref{eq:vartheta_map} to that of a system coupled to a quantized harmonic oscillator mode, and we can borrow some terminology. Eq.~\eqref{eq:vartheta_map} describes a coupled system of two modes consisting of qubit and drive, with the pair of quantum numbers $(m,\alpha)$ for the drive and the qubit, respectively. Here, $\alpha$ is an index for the qubit system eigenstates in both the original and expanded Hilbert spaces, and $m$ represents the ``photon number'' of drive photons that get added or subtracted.
When the drive amplitude is zero, the eigenstates of $\widetilde{H}_F$ are product states $\ket{\widetilde{u}_{m,\alpha}} =\ket{m}\otimes\ket{\alpha}$, with eigenenergies $\varepsilon_{m,\alpha}=E_\alpha+m\omega$, where $E_\alpha$ are the energies of the physical states $\ket{\alpha}$ in the original Hilbert space. When the drive amplitude is nonzero, the eigenstates $\ket{\widetilde{u}_{m,\alpha}} =\ket{m,\alpha}$ of $\widetilde{H}_F$ are no longer product states due to the hybridization between the qubit and the drive, and the corresponding eigenenergies are the aforementioned quasienergies $\varepsilon_{m,\alpha}$.

\subsection{Generalized Floquet theory} \label{app:generalized_Floquet}
We can further generalize the extended Hilbert space representation to a scenario where $\hat{H}(t)$ is not strictly periodic but has a quasi-periodic structure.
Suppose the Hamiltonian $\hat{H}(t)$ can be written in the form $\hat{H}(\theta,t)$, where the periodicity is with respect to $\theta$, i.e., $\hat{H}(\theta,t)=\hat{H}(\theta+2\pi,t)$. Similar analysis as in Appendix~\ref{sec:extended_HS} can be done and the Floquet Hamiltonian will take the following form
\begin{equation}
    \widetilde{H}_F(t) = \hat{H}(\hat{\vartheta},t) + \dot{\theta}\hat{m} ,
\end{equation}
where $\dot{\theta} = \text{d}\theta/\text{d}t$. Written in the $\hat{m}$-basis explicitly, we have
\begin{equation} \label{eq:K_1qb_compact}
    \widetilde{H}_F(t) = \sum_{n,k=-\infty}^{+\infty} \ket{n+k}\!\bra{n} \otimes \hat{H}^{(k)}(t) + \sum_{k=-\infty}^{+\infty}k\ket{k}\!\bra{k} \otimes \dot{\theta}I ,
\end{equation}
where $\hat{H}^{(k)}(t)$ are the coefficients in the Fourier decomposition of the Hamiltonian, i.e., $\hat{H}(\theta, t) = \displaystyle\sum_{k=-\infty}^{+\infty} \hat{H}^{(k)}(t) e^{ik\theta}$.

We note that the Floquet Hamiltonian $\widetilde{H}_F$ presented in Appendix~\ref{sec:extended_HS} is time-independent; whereas the Floquet Hamiltonian $\widetilde{H}_F(t)$ considered here retains explicit time dependence. This residual time dependence arises from variables that evolve in time but are not encapsulated within the fast-varying $\theta$. When these additional variables vary slowly compared to the rapid change in $\theta$, the construction of the Floquet Hamiltonian effectively separates the system's dynamics into fast and slowly varying components. In our implementation, this separation is both valid and useful, as the remaining time-dependent parameters evolve on a much slower timescale than $\theta$, and our objective is to design these slowly varying components.

This generalization leads to the capability of treating more complicated systems with multiple incommensurate quasi-periodicities. For example, consider a multi-qubit system with multiple drives:
\begin{equation}
    \hat{H}(t) = \sum_i \hat{H}_{0i} + \Omega_i\cos(\theta_i)\hat{n}_i.
\end{equation}
In this case, the Floquet Hamiltonian will take the following form
\begin{equation}
    \widetilde{H}_F(t) = \hat{H}(\{\hat{\vartheta}_i\},t) + \sum_i \dot{\theta_i}\hat{m}_i.
\end{equation}
Formulated this way, it is then possible to co-design multiple gates in the same framework.

\section{Unitary propagator} \label{app:unitary_propagator}
We discuss the unitary propagator during the gates for both adiabatic and nonadiabatic gates leveraging the extended Hilbert space representation $\mathcal{K}$.
\subsubsection{Adiabatic Gates}
In adiabatic gates, the key is to keep the state in the intended instantaneous Floquet mode as much as possible. At $t=0$, we prepare the initial state $\ket{u(0)}=\ket{\alpha}$ in $\mathcal{H}$, which maps to $\ket{\widetilde{u}_{m,\alpha}(0)}=\ket{m}\otimes\ket{\alpha}$ in $\mathcal{K}$. When $0<t<t_g$, where $t_g$ is the total gate duration, we can derive the evolution propagator according to the standard adiabatic theorem, assuming that there are no quasienergy degeneracies and the control parameters vary slowly, $\lambda(t) = [\Omega(t),\ \omega_\text{inst}(t)]^T$. The desired propagator in $\mathcal{K}$ for $0\leq t\leq t_g$ can be written as
\begin{equation}
    \widetilde{U}(t) = \sum_{m,\alpha} e^{-i\int_0^t\varepsilon_{m,\alpha}(\lambda(t))\text{d}t}\ket{\widetilde{u}_{m,\alpha}(\lambda(t))}\!\bra{\widetilde{u}_{m,\alpha}(\lambda(0))},
\end{equation}
where we choose the gauge condition $\braket{\widetilde{u}_{m,\alpha(\lambda(t))}|\nabla_\lambda \widetilde{u}_{m,\alpha}(\lambda(t))}\cdot\dot{\lambda}(t)=0$ to account for geometric phases during the evolution~\cite{Sakurai_Napolitano_2020}. At $t=t_{g}$, we have $\ket{\widetilde{u}_{m,\alpha}(t_{g})}=U(t_{g})\ket{\widetilde{u}_{m,\alpha}(0)}$. Finally, we map back to $\mathcal{H}$ by $\ket{u(t_g)} = \braket{\theta(t_g)|\widetilde{u}_{m,\alpha}(t_{g})}$, where $\bra{\theta(t_g)}=\sum_m e^{im\theta(t_g)}\bra{m}$.

\subsubsection{Nonadiabatic Gates}
In the nonadiabatic gate approach, we no longer require the quantum state to remain in the instantaneous eigenstate of the system throughout its evolution. Instead, we intentionally engineer nonadiabatic transitions during the gate, especially the frequency modulation, to steer the system toward a desired final state. The key idea is that even if the state departs from its initial eigenstate during the gate, the overall evolution can still yield a high-fidelity operation, provided the final state aligns with the target. At $t=0$, we prepare the initial state $\ket{u(0)} = \ket{\alpha}$ in $\mathcal{H}$, which corresponds to $\ket{\widetilde{u}_{m,\alpha}(0)} = \ket{m} \otimes \ket{\alpha}$ in $\mathcal{K}$. The goal is to design the time-dependent control vector $\lambda(t) = [\Omega(t),\ \omega_\text{inst}(t)]^T$ such that the system's unitary propagator at $t=t_{g}$ in $\mathcal{K}$ takes the desired form
\begin{equation} \label{eq:u_nonaidabatic}
    \widetilde{U}(t_{g}) = \sum_{m,\alpha} e^{-i\phi_{m',\alpha'}(\lambda(t_{g}))} \ket{\widetilde{u}_{m',\alpha'}(\lambda(t_{g}))}\!\bra{\widetilde{u}_{m,\alpha}(\lambda(0))},
\end{equation}
where $\ket{\widetilde{u}_{m',\alpha'}(\lambda(t_{g_-}))}$ is the desired final Floquet mode evolved from the initial mode $\ket{\widetilde{u}_{m,\alpha}(\lambda(0))}$, and $\phi_{m',\alpha'}$ is the accumulated phase and $\{m',\alpha'\}\neq\{m,\alpha\}$. Note that Eq.~\eqref{eq:u_nonaidabatic} generally only holds at $t=t_g$. Finally, the state is projected back to $\mathcal{H}$.

\section{Fast quasiadiabatic dynamics} \label{app:faquad}
We adopt the fast quasiadiabatic (FAQUAD) protocol and its generalized version~\cite{martinez-garaot_Fast_2015,wu_Adiabaticity_2020,garcia-ripoll_Quantum_2020a,guery-odelin_Shortcuts_2019} to accelerate adiabatic dynamics by shaping the temporal profile of the control parameter according to the instantaneous energy gap. FAQUAD redistributes the rate of change in the control parameter to slow down in regions where the gap between eigenstates is small (e.g., near avoided crossings) and speed up where the gap is large. This results in a time-dependent trajectory that maintains homogeneous adiabaticity while reducing total gate duration. 

Specifically, in a simplified two-level example with Hamiltonian \( \hat{H}(t)=\hat{H}[\lambda(t)] \) where \(\lambda(t)\) is the control parameter, FAQUAD imposes the condition
\begin{equation} \label{eq:FAQUAD}
    \left|\frac{\braket{\phi_1(t)|\partial_t \phi_2(t)}}{E_1(t)-E_2(t)}\right| = \left|\frac{\bra{\phi_1(t)}\partial_t \hat{H} \ket{\phi_2(t)}}{(E_1(t)-E_2(t))^2}\right| = c,
\end{equation}
where \( c \) is a constant set by the desired adiabaticity level, \(\phi_i(t)\) and \(E_i(t)\) are instantaneous eigenvectors and eigenenergies of \(\hat{H}(t)\).
Following Eq.~\eqref{eq:FAQUAD}, applying the chain rule yields
\begin{equation} \label{eq:ODE_lambda_t}
    \frac{\text{d}\lambda}{\text{d}t} = \pm c \left| \frac{E_1(\lambda) - E_2(\lambda)}{\braket{\phi_1(\lambda)| \partial_\lambda\phi_2(\lambda)}} \right|=\pm c \left| \frac{(E_1(\lambda) - E_2(\lambda))^2}{\bra{\phi_1(\lambda)} \partial_\lambda \hat{H} \ket{\phi_2(\lambda)}} \right|,
\end{equation}
where $\pm$ indicates the monotonous increase or decrease of $\lambda(t)$. The differential equation in Eq.~\eqref{eq:ODE_lambda_t} can be solved for \( \lambda(t) \) given the boundary conditions $\lambda(0)$ and $\lambda(t_g)$ where $t_g$ is the total duration of the specified process. 

In practice, to ensure that the designed control parameter \(\lambda(t)\) is smooth at both the beginning and the end of the protocol, thus complying with hardware constraints such as finite bandwidth, we introduce a smoothness envelope function \(s(t)\) that is multiplied on the right-hand side of Eq.~\eqref{eq:ODE_lambda_t}. An example choice for \(s(t)\) is
\begin{equation}
s(t) = 
\begin{cases}
\begin{array}{@{}l@{}}
\displaystyle \frac{1}{2}\left(1 - \cos\left(\frac{\pi t}{rt_g}\right)\right) \\
\text{\hspace*{3.3cm}} \text{for } 0 \leq t < rt_g
\end{array} \\[8pt]

\begin{array}{@{}l@{}}
1 \\
\text{\hspace*{3.3cm}}\text{for } rt_g \leq t < (1-r)t_g \text{\hspace*{0.1cm}},
\end{array} \\[8pt]

\begin{array}{@{}l@{}}
\displaystyle \frac{1}{2}\left(1 + \cos\left(\frac{\pi (t - (1-r)t_g)}{rt_g}\right)\right) \\
\text{\hspace*{3.3cm}} \text{for } (1-r)t_g \leq t \leq t_g
\end{array}
\end{cases}
\end{equation}
where $0<r<1/2$ determines the duration of the initial and final regions that we would like to enforce smoothness. In this work, we have used $r=0.05$ in all examples to obtain a sufficiently smooth pulse shape without compromising the gate speed. Then, instead of Eq.~\eqref{eq:ODE_lambda_t}, we solve 
\begin{equation} \label{eq:ODE_lambda_t_smooth}
    \frac{\text{d}\lambda}{\text{d}t} =\pm c \left| \frac{(E_1(\lambda) - E_2(\lambda))^2}{\bra{\phi_1(\lambda)} \partial_\lambda \hat{H} \ket{\phi_2(\lambda)}} \right|s(t)
\end{equation}
for $\lambda(t)$. This choice ensures that \(\dot{\lambda}(t)\) vanishes at both endpoints, \(t = 0\) and \(t = t_g\), resulting in a smooth \(\lambda(t)\) at boundary points.

A generalized multi-level extension of the FAQUAD method can be defined to account for the adiabatic evolution of multiple pairs of instantaneous eigenstates. Let \( Q = \{(p_i, q_i)\} \) be a set of index pairs denoting the relevant transitions between instantaneous eigenstates \(\ket{\phi_{p_i}(t)}\) and \(\ket{\phi_{q_i}(t)}\). Let \( W = \{w_i\} \) be a corresponding set of positive weights that assign relative importance to each pair in \(Q\). The adiabatic constraint is then imposed via
\begin{equation} \label{eq:FAQUAD_multi}
    c = \max_i \left\{ w_i \left| \frac{ \bra{\phi_{p_i}(t)} \partial_t \hat{H}(t) \ket{\phi_{q_i}(t)} }{ ( E_{p_i}(t) - E_{q_i}(t) )^2 } \right| \right\}.
\end{equation}
For instance, in a three-level system where adiabaticity between the pairs \((1,2)\) and \((1,3)\) are emphasized equally, we obtain
\begin{equation} \label{eq:FAQUAD_multi_example}
    c = \max \bigg\{\left|\frac{\bra{\phi_1(t)}\partial_t \hat{H} \ket{\phi_2(t)}}{(E_1(t)-E_2(t))^2}\right|, \left|\frac{\bra{\phi_1(t)}\partial_t \hat{H} \ket{\phi_3(t)}}{(E_1(t)-E_3(t))^2}\right|\bigg\}.
\end{equation}
$\lambda(t)$ can be solved similarly as in the original FAQUAD method.
This generalization enables flexible control strategies in multi-level quantum systems by tailoring the adiabatic condition to prioritize the most critical adiabaticity while still accounting for others.

\section{Simulation details} \label{app:sim_details}

\subsection{Definitions} \label{app:definitions}
We use QuTiP~\cite{lambert_QuTiP_2024} to perform the time-dependent gate simulations in this work, and use \texttt{jaxquantum}~\cite{jha2024jaxquantum} for the construction and diagonalization of the Floquet Hamiltonians. In order to efficiently track the dressed eigenstates of the Floquet Hamiltonian as we vary the control parameters, we utilize the state labeling procedure developed in Ref.~\cite{chowdhury2025theoryquasiparticlegenerationmicrowave}. The average gate fidelity $\mathcal{F}$ is computed as~\cite{pedersen_Fidelity_2007}:
\begin{equation}
    \mathcal{F} = \frac{\mathrm{Tr}(\hat{M}^\dag \hat{M}) + |\mathrm{Tr}(\hat{U}_\text{ideal}^\dag \hat{M})|^2}{n(n+1)},
\end{equation}
where $\hat{M}$ is the system propagator of the designed gate in the computational basis, $\hat{U}_\text{ideal}$ is the ideal unitary propagator of the desired gate, and $n$ is the dimension of the subspace. $\hat{M}$ is not necessarily unitary; it is constructed by evolving each computational basis state under the full system dynamics and projecting the resulting states back onto the computational basis, so that any leakage outside the computational subspace is reflected in its nonunitarity. To avoid confusion, we report control error $1-\mathcal{F}$ in this paper.

\subsection{CZ gate} \label{app:conventional_CZ}
\begin{figure*}[htbp]
    \centering
    \includegraphics[width=0.99\linewidth]{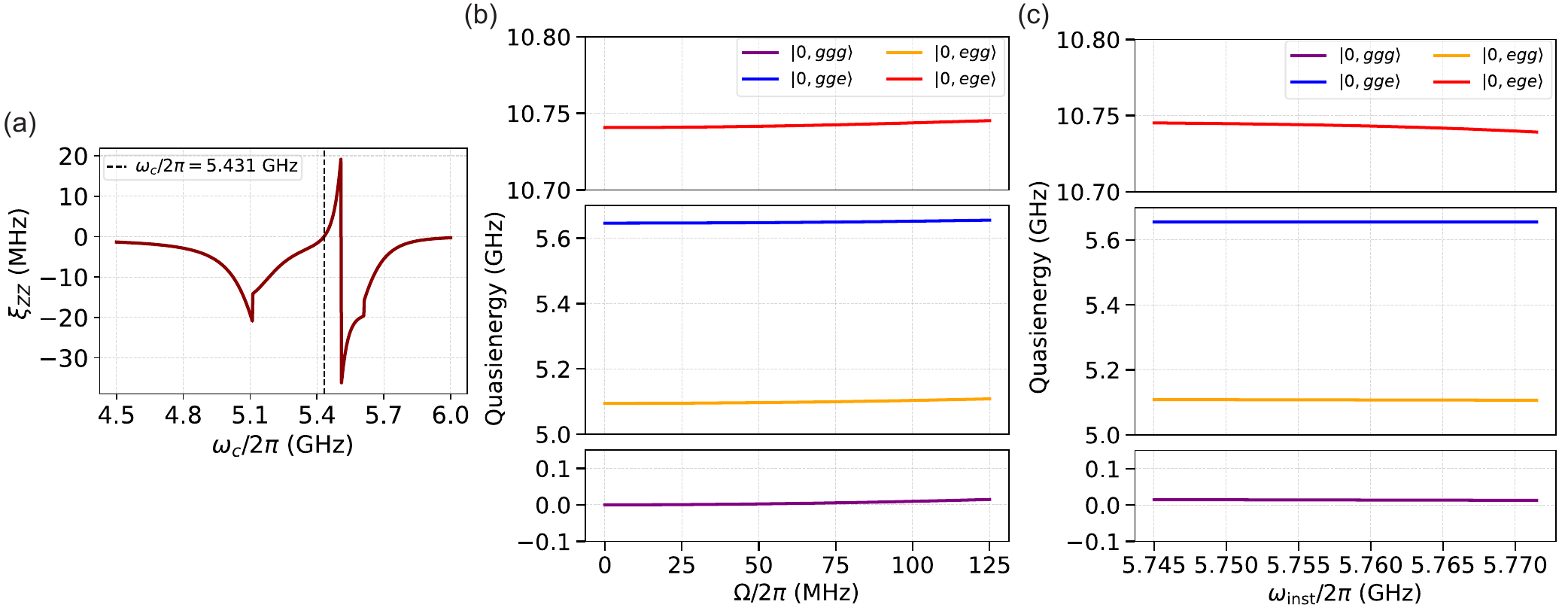}
    \caption{CZ gate. (a) Static ZZ interaction as a function of coupler frequency $\omega_c$. We choose $\omega_c/2\pi=5.431$ GHz to have a vanishing ZZ interaction. (b)(c) Quasienergy spectrum of the computational Floquet modes $\ket{0,ggg},\ket{0,gge},\ket{0,egg},\ket{0,ege}$ as a function of drive amplitude $\Omega(t)$ and instantaneous frequency $\omega_\text{inst}(t)$, respectively.}
    \label{fig:cz_app}
\end{figure*}
\begin{table}[h]
\renewcommand{\arraystretch}{1.7}
\centering
\begin{tabular}{|c|c|c|}
\hline
$\omega_a/2\pi$ & $\omega_b/2\pi$ & $\omega_c/2\pi$  \\
\hline
5.111 GHz & 5.612 GHz & 5.431 GHz  \\
\hline
\hline
$\alpha_a/2\pi$ & $\alpha_b/2\pi$ & $\alpha_c/2\pi$ \\
\hline
-231.5 MHz & -249.9 MHz & -294.7 MHz \\
\hline
\hline
$g_{ac}/2\pi$ & $g_{bc}/2\pi$ & $g_{ab}/2\pi$ \\
\hline
75.2 MHz & 82.5 MHz & 7.2 MHz \\
\hline
\end{tabular}
\caption{Device parameters for the CZ gate.}
\label{tab:CZ_gate_params}
\end{table}
The device parameters we use to simulate the example in Section~\ref{sec:traditional_CZ} are summarized in Table~\ref{tab:CZ_gate_params}. Fig.~\ref{fig:cz_app}(a) shows the static ZZ interaction as a function of $\omega_c$, which is picked to be $\omega_c/2\pi=5.431$ GHz so that the ZZ interaction vanishes when static. Figs.~\ref{fig:cz_app}(b) and~\ref{fig:cz_app}(c) show the quasienergy spectrum of the computational Floquet modes $\ket{0,ggg},\ket{0,gge},\ket{0,egg},\ket{0,ege}$ as a function of $\Omega(t)$ and $\omega_\text{inst}(t)$, respectively. We use these to calculate the ZZ interaction in Figs.~\ref{fig:cz_gate}(a) and~\ref{fig:cz_gate}(c). It is worth noting that the ZZ interaction during the amplitude ramp up decreases despite a slight increase in the quasienergy of $\ket{0,ege}$. This occurs because the other three computational Floquet modes also experience an upward shift in their quasienergies. In contrast, during the frequency modulation, the bending down of the $\ket{0,ege}$ quasienergy leads to a significant reduction in the ZZ interaction. The maximum ZZ interaction in this example is roughly $-9.61$ MHz, which restricts the CZ gate duration to be at least over 100 ns in a adiabatic gate by rough estimation. The key in finding a fast CZ gate is to be able to identify a large ZZ interaction region; the caveat is that the Floquet modes may come closer to one another, leading to more leakage error.

\subsection{Always-on CZ gate} \label{app:alwayson_CZ}
\begin{figure*}[htbp]
    \centering
    \includegraphics[width=0.99\linewidth]{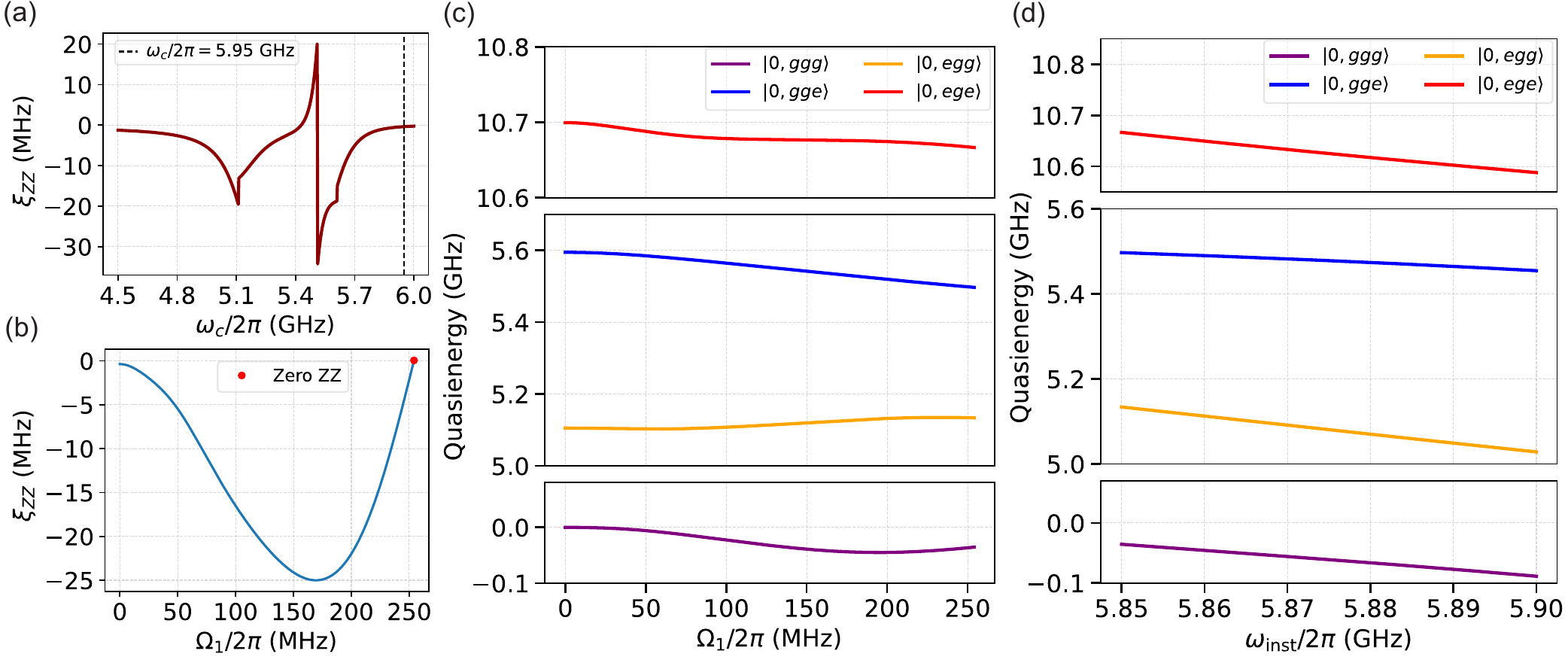}
    \caption{Always-on CZ gate. (a) Static ZZ interaction (without drive) as a function of coupler frequency $\omega_c$. We choose $\omega_c/2\pi=5.95$ GHz to have a small nonzero ZZ interaction. (b) Idling ZZ interaction (with drive) as a function of drive amplitude $\Omega_1$ when instantaneous frequency is fixed at $\omega_\text{inst}(t)/2\pi=\omega_0/2\pi=5.85$ GHz. Red dot at $\Omega_1/2\pi=254.1$ MHz is where we operate with vanishing idling ZZ interaction. (c)(d) Quasienergy spectrum of the computational subspace consisting of Floquet modes $\ket{0,ggg},\ket{0,gge},\ket{0,egg},\ket{0,ege}$ as a function of drive amplitude $\Omega_1$ and instantaneous frequency $\omega_\text{inst}(t)$, respectively. Note that in the always-on CZ gate the drive amplitude remains a constant, but (c) indicates an adiabatic mapping between the Floquet modes and bare eigenstates of the undriven system.}
    \label{fig:alwayson_cz_app}
\end{figure*}
\begin{table}[h]
\renewcommand{\arraystretch}{1.7}
\centering
\begin{tabular}{|c|c|c|}
\hline
$\omega_a/2\pi$ & $\omega_b/2\pi$ & $\omega_c/2\pi$  \\
\hline
5.111 GHz & 5.612 GHz & 5.950 GHz  \\
\hline
\hline
$\alpha_a/2\pi$ & $\alpha_b/2\pi$ & $\alpha_c/2\pi$ \\
\hline
-231.5 MHz & -249.9 MHz & -298.7 MHz \\
\hline
\hline
$g_{ac}/2\pi$ & $g_{bc}/2\pi$ & $g_{ab}/2\pi$ \\
\hline
72.3 MHz & 79.2 MHz & 7.2 MHz \\
\hline
\end{tabular}
\caption{Device parameters for the always-on CZ gate.}
\label{tab:alwayson_CZ_gate_params}
\end{table}
The device parameters we use to simulate the example in Section~\ref{sec:always_on_CZ} are summarized in Table~\ref{tab:alwayson_CZ_gate_params}. Fig.~\ref{fig:alwayson_cz_app}(a) shows the static ZZ interaction (without drive) as a function of coupler frequency $\omega_c$. We choose $\omega_c=5.95$ GHz for the static ZZ interaction to be nonzero. We then choose an appropriate set of always-on drive amplitude $\Omega_1$ and frequency $\omega_0$ such that the idling ZZ interaction is zero. Fig.~\ref{fig:alwayson_cz_app}(b) shows the idling ZZ interaction as a function of $\Omega_1$ when $\omega_0/2\pi=5.85$ GHz. The red dot indicates that the idling ZZ vanishes at $\Omega_1 = 254.1$~MHz. We emphasize that this choice of device and control parameters is not unique. With the fixed device parameters in Table~\ref{tab:alwayson_CZ_gate_params}, different control parameter sets can meet the basic requirements for implementing a CZ gate. However, different choices may yield different gate fidelities and durations due to variations in ZZ interaction strength and spectral structure. Fig.~\ref{fig:alwayson_cz_app}(c) shows the quasienergy spectrum of the computational basis states consisting of Floquet modes $\ket{0,ggg},\ket{0,gge},\ket{0,egg},\ket{0,ege}$ as a function of $\Omega_1$. Although the drive amplitude remains constant during the always-on CZ gate, we require an adiabatic mapping between the computational states and the bare eigenstates of the undriven system to enable proper readout. Fig.~\ref{fig:alwayson_cz_app}(d) shows the quasienergy spectrum as a function of $\omega_\text{inst}(t)$.

\subsection{Z gate} \label{app:z_gate}
\begin{figure}[htbp]
    \centering
    \includegraphics[width=0.99\linewidth]{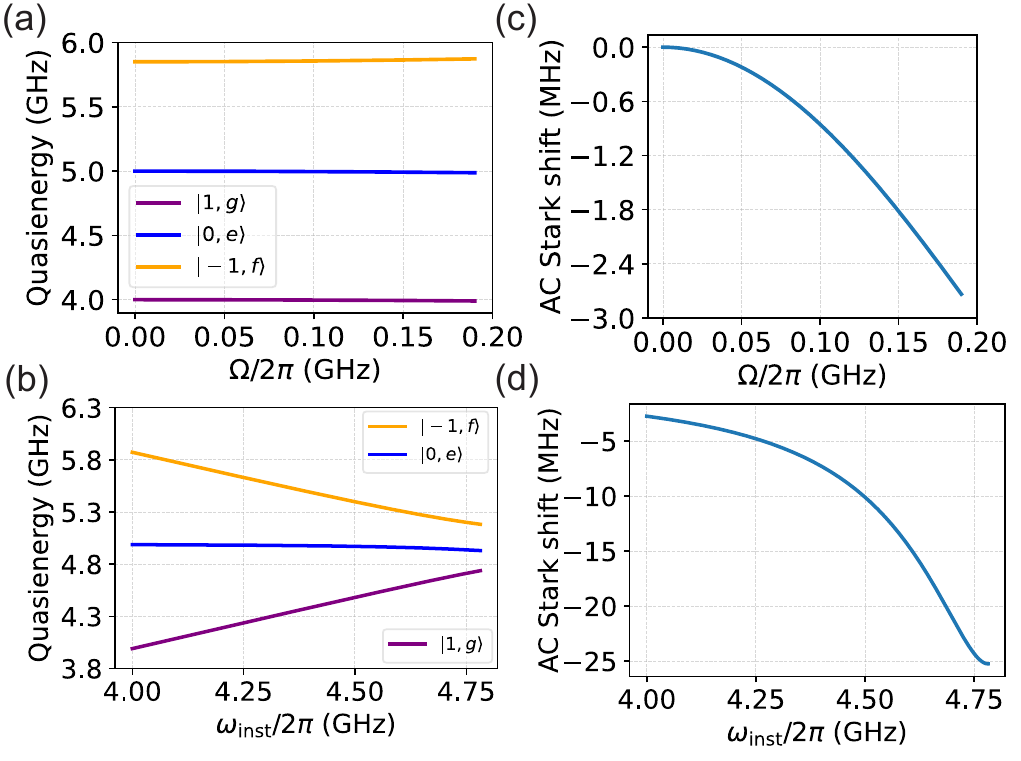}
    \caption{Z gate. (a)(b) Quasienergy spectrum of the Floquet modes $\{\ket{1,g},\ket{0,e},\ket{-1,f}\}$-subspace as a function of $\Omega(t)$ and $\omega_\text{inst}(t)$. We show $\omega_\text{inst}(t)$ up to $\omega^*$. (c)(d) AC Stark shift as a function of $\Omega(t)$ and $\omega_\text{inst}(t)$.}
    \label{fig:z_app}
\end{figure}
\begin{table}[h!]
\renewcommand{\arraystretch}{1.7}
\centering
\begin{tabular}{|c|c|c|c|}
\hline
$\omega_q/2\pi$ & $\alpha/2\pi$ & $\Omega_1/2\pi$ & $[\omega_0,\omega^*]/2\pi$ \\
\hline
5 GHz & -150 MHz & $190$ MHz & $[4.0,4.78]$ GHz \\
\hline
\end{tabular}
\caption{Device and control parameters for the Z gate.}
\label{tab:Z_gate_params}
\end{table}

We show additional details about the Z gate. The device parameters we use to simulate the example in Section~\ref{sec:Z_gate} are summarized in Table~\ref{tab:Z_gate_params}. Figs.~\ref{fig:z_app}(a) and~\ref{fig:z_app}(b) show the quasienergy spectrum as a function of $\Omega(t)$ and $\omega_\text{inst}(t)$ respectively. Fig.~\ref{fig:z_app}(a) shows a very separated spectrum because the initial drive frequency $\omega_0$ is further off-resonant with the qubit frequency $\omega_q$. Therefore, the amplitude ramp up and down is even faster (under 4 ns) for comparable drive amplitudes. There is some flexibility in adjusting $\omega_0$ to balance the duration of the amplitude and frequency modulation. The exact avoided crossing between $\ket{1,g}$ and $\ket{0,e}$ is found to be $\omega^*/2\pi=4.78$ GHz. We restrict $\omega_1 \leq \omega^*$ to maintain the adiabaticity during the frequency modulation.
As shown in Figs.~\ref{fig:z_app}(c) and~\ref{fig:z_app}(d), the AC Stark shift is weak during the amplitude modulation. In contrast, the spectrum changes significantly with $\omega_\text{inst}(t)$, leading to a stronger AC Stark shift during the frequency modulation.

\subsection{X gate} \label{app:x_y_gate}
\begin{figure}[htbp]
    \centering
    \includegraphics[width=0.99\linewidth]{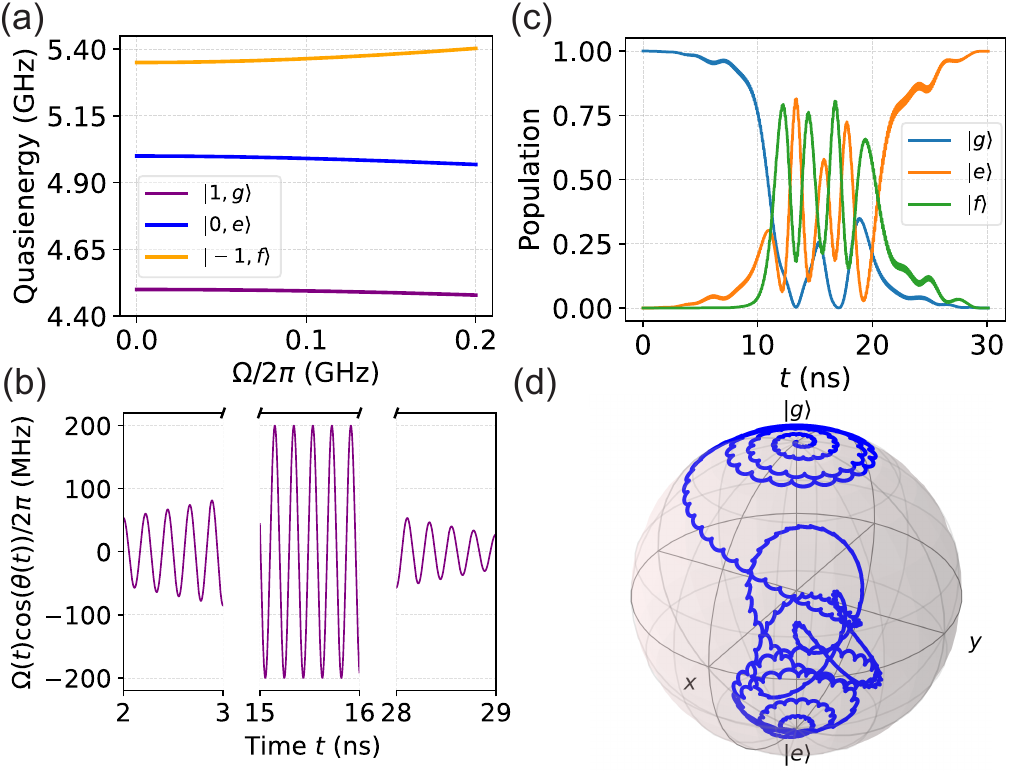}
    \caption{X gate. (a) Quasienergy spectrum of the Floquet modes $\{\ket{1,g},\ket{0,e},\ket{-1,f}\}$-subspace as a function of $\Omega(t)$. (b) Segments of the full pulse $\Omega(t)\cos{\theta(t)}$. (c) An example of the population overlap evolution in $\ket{g},\ket{e},\ket{f}$ during the X gate if initialized in $\ket{g}$. (d) Trajectory of the state evolution of the X gate on the Bloch sphere.}
    \label{fig:x_y_app}
\end{figure}

\begin{table}[h!]
\renewcommand{\arraystretch}{1.7}
\centering
\begin{tabular}{|c|c|c|c|}
\hline
$\omega_q/2\pi$ & $\alpha/2\pi$ & $\Omega_1/2\pi$ & $\omega_0/2\pi$ \\
\hline
5 GHz & -150 MHz & $200$ MHz & $4.5$ GHz \\
\hline
\end{tabular}
\caption{Device and control parameters for the X gate.}
\label{tab:X_gate_params}
\end{table}

We show more simulation details about the X gate. The device parameters we use to simulate the example in Section~\ref{sec:XY_gate} are summarized in Table~\ref{tab:X_gate_params}. The simple parameterized model we use for $\omega_\text{inst}(t)$ is
\begin{equation} \label{eq:cosine_hold}
\omega_\text{inst}(t) =
\begin{cases}
\begin{array}{@{}l@{}}
\displaystyle \frac{\omega_0 + \omega_1}{2}
+ \frac{\omega_0 - \omega_1}{2}
\cos\left( \frac{\pi (t - t_a)}{t_\omega} \right) \\
\text{\hspace*{1.3cm}} \text{for } t_a \leq t \leq t_a + t_\omega
\end{array} \\[8pt]

\begin{array}{@{}l@{}}
\omega_1 \\
\text{\hspace*{1.3cm}}\text{for } t_a + t_\omega < t \leq t_a + t_\omega + t_h\text{\hspace*{1.1cm}}
\end{array} \\[8pt]

\begin{array}{@{}l@{}}
\displaystyle \frac{\omega_0 + \omega_1}{2}
+ \frac{\omega_1 - \omega_0}{2}
\cos\left( \frac{\pi (t - t_a - t_\omega - t_h)}{t_\omega} \right) \\
\text{\hspace*{1.3cm}} \text{for } t_a + t_\omega + t_h < t \leq t_a + 2t_\omega + t_h
\end{array}
\end{cases}
\end{equation}
Fig.~\ref{fig:x_y_app}(a) shows the quasienergy spectrum as a function of $\Omega(t)$, which are considerably separated so that we can ramp up and down the amplitude in 6 ns without introducing much nonadiabatic error. Segments of the full pulse $\Omega(t)\cos[{\theta(t)}]$ are shown in Fig.~\ref{fig:x_y_app}(b). It is easy to see the change of amplitude and instantaneous frequency during the beginning, middle, and final stages of the pulse. To illustrate what the evolution during the X gate looks like, we prepare the initial state in $\ket{g}$ and plot the population overlap in $\ket{g},\ket{e},\ket{f}$ in Fig.~\ref{fig:x_y_app}(c). We also show the trajectory on the Bloch sphere in Fig.~\ref{fig:x_y_app}(d). As expected, the state leaks outside the computational subspace during the middle of the gate but eventually mostly ends up in the desired final state $\ket{e}$. We also show the trajectory of the state evolution the X gate on the initial state $\ket{0}$ in the rotating frame. In this example, we restrict the optimization of $\omega_\text{inst}(t)$ to a subspace of the cosine-and-hold pulse shape, as the resulting control error is comparable to that of state-of-the-art X gates in experiments. Further improvement of control error can be done by enlarging the optimization subspace with more degrees of freedom.

\subsection{Hadamard gate} \label{app:h_gate}
\begin{figure}
    \centering
    \includegraphics[width=0.99\linewidth]{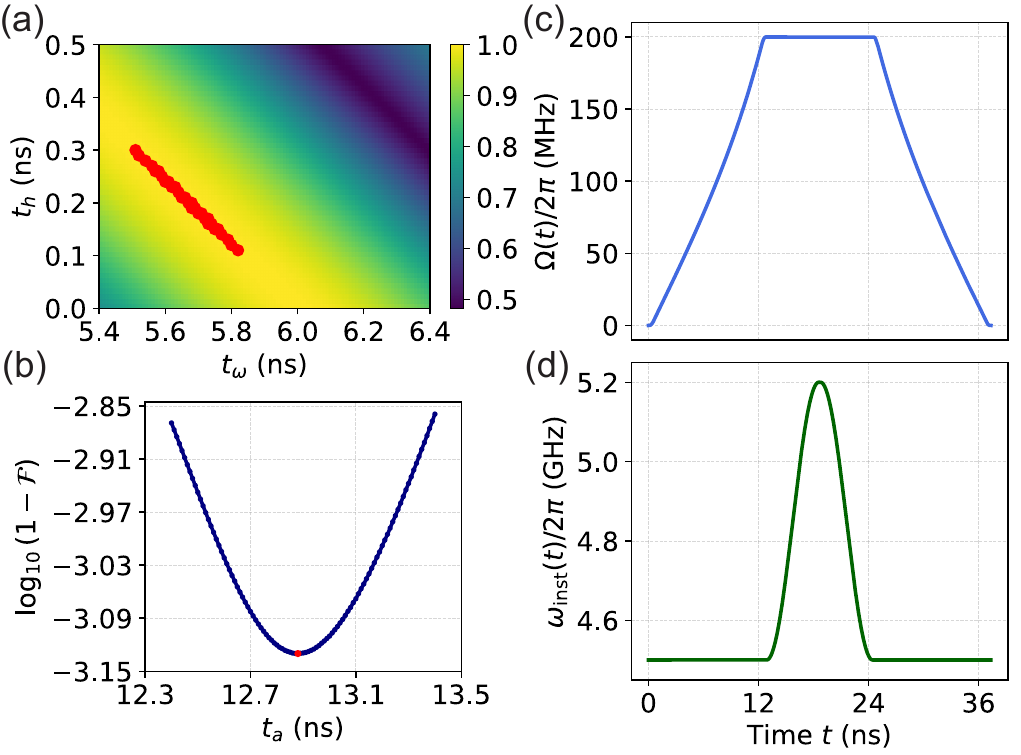}
    \caption{Hadamard gate. (a) Average overlap as a function of $t_\omega,t_h$ when $\omega_1/2\pi=5.2$ GHz at the end of the frequency modulation during the Hadamard gate. Red area in the middle indicates top candidates for potential high-fidelity Hadamard gates. (b) The Hadamard gate control error as a function of $t_a$ for one specific set of $t_\omega,t_h,\omega_1$ among the top candidates in (a). (c)(d) The optimized control waveforms $\Omega(t)$ and $\omega_\text{inst}(t)$ for the best Hadamard gate.}
    \label{fig:h_gate}
\end{figure}
We implement the Hadamard gate based on the nonadiabatic gate principles. The simulated operation has the following matrix representation:
\begin{equation}
    \mathrm{H}_{\phi_1,\phi_2,\phi_3} = \displaystyle e^{i\phi_1} \begin{bmatrix} 1 & 0 \\ 0 & e^{i\phi_2} \end{bmatrix} \frac{1}{\sqrt{2}} \begin{bmatrix} 1 & 1 \\ 1 & -1 \end{bmatrix} \begin{bmatrix} 1 & 0 \\ 0 & e^{i\phi_3} \end{bmatrix}.
\end{equation}

We consider the same system, with device and control parameters identical to those summarized in Table~\ref{tab:X_gate_params} for the X gate in Section~\ref{sec:XY_gate}. The gate principle of the Hadamard gate is highly similar to that of the X gate, except for the change of the goal during the frequency modulation. Here, the objective of frequency modulation is to drive a partial (50\%) population transfer, thereby creating an equal superposition of the two Floquet modes, $\ket{1,g}$ and $\ket{0,e}$, regardless of the initial Floquet mode, while simultaneously minimizing leakage to $\ket{-1,f}$. The design of control parameters $t_a,\omega_1,t_\omega,t_h$ as well as $\Omega(t)$ and $\omega_\text{inst}(t)$ follows the same procedure as in Section~\ref{sec:XY_gate}. Fig.~\ref{fig:h_gate}(a) shows the average overlap as a function of $t_\omega,t_h$ when $\omega_1/2\pi=5.2$ GHz near the optimal region. We then pick the top candidates within the region (indicated by the red area) and simulate the Hadamard gate as a function of different $t_a$ and its corresponding $\Omega(t)$, as shown in Fig.~\ref{fig:h_gate}(b). The best Hadamard gate we find has control error of approximately $0.07\%$ and a total gate duration of $37.4$ ns. The optimized control waveforms $\Omega(t)$ and $\omega_\text{inst}(t)$ are shown in Figs.~\ref{fig:h_gate}(c) and~\ref{fig:h_gate}(d).

\section{Discussion on hardware implementation} \label{app:hardware}
We discuss the hardware requirements necessary to implement the control pulses proposed in this work, focusing in particular on bandwidth and sampling rate. In superconducting qubit platforms, microwave control pulses are typically synthesized using single-sideband (SSB) modulation. Direct digital sythesis may be a more flexible choice if possible. For the pulses proposed in this work, we show that they can be readily implemented using the same SSB modulation technique without requiring significant modifications to existing experimental setups.

We use the pulses designed in Section~\ref{sec:traditional_CZ} as a concrete example. The eventual pulse we would like to generate is $\Omega(t)\cos{\theta(t)}$, where $\theta(t)=\int_0^{t_g}\omega_\text{inst}(t)$. The profiles of $\Omega(t)$ and $\omega_\text{inst}(t)$ are depicted in Fig.~\ref{fig:cz_gate}(f). The ranges for $\Omega(t)/2\pi$ and $\omega_\text{inst}(t)/2\pi$ are $[0,125]$ MHz and $[5.745,5.7615]$ GHz, respectively. We can specify the RF tone to be, for instance, $\omega_\text{RF}/2\pi=\omega_0/2\pi=5.745$ GHz. Then the baseband pulse will be $\displaystyle I(t) = \Omega(t)\cos{(\theta(t)-\omega_\text{RF}t)} = \Omega(t)\cos{\bigg(\int_0^{t_g}\big(\omega_\text{inst}(t)-\omega_\text{RF}\big)t\bigg)}$, with the instantaneous frequency in the range of $[0,16.5]$ MHz. We then estimate the bandwidth of $x(t)$ to be less than $60$ MHz using a $-40$ dB cutoff. Therefore, a standard off-the-shelf AWG (e.g., Keysight M3202A with 1 GSa/s sampling rate and 400 MHz bandwidth) is expected to generate this pulse accurately.

For the pulses designed in Section~\ref{sec:nonadiabatic_gates} and Appendix~\ref{app:h_gate}, a more sophisticated AWG may be needed (e.g., Keysight M5300A with 4.8 GSa/s sampling rate and 2 GHz bandwidth). This is mainly because the frequency modulation range is significantly larger—exceeding 700 MHz—resulting in an baseband pulse bandwidth estimated at approximately 800 MHz using a \(-40\) dB cutoff. We note that such a large frequency modulation range is not necessary to implement single-qubit gates, whose requirements are generally less stringent. The frequency modulation range can be reduced to lower the required bandwidth and alleviate hardware demands while maintaining desired gate fidelity.

% \clearpage
% \newpage
% \bibliographystyle{apsrev4-2}
\bibliography{references}

\end{document}